\documentclass[letter]{aa}  

\usepackage{graphicx}
\usepackage{txfonts}
\usepackage{comment}
\usepackage{color}
\usepackage{xcolor}
\usepackage{url}
\usepackage{hyperref}
\usepackage{multirow}
\usepackage{float}
\usepackage{orcidlink}
\usepackage{soul} 

\definecolor{myblue}{HTML}{6a359c} 
\definecolor{myviolet}{HTML}{756BB1}
\definecolor{mygreen}{HTML}{7FCDBB}
\definecolor{myyellow}{HTML}{FFFDD1}

\hypersetup{
	colorlinks,
	linkcolor={myblue},
	citecolor={myblue},
	urlcolor={myblue}}

\newcommand{\ME}{\rm{M}$_{\oplus}$} 
\newcommand{\RE}{\rm{R}$_{\oplus} \,$} 
\newcommand{\Mearth}{\text{M}_{\oplus}} 
\newcommand{\Msun}{\text{M}_{\odot} }
\newcommand{\Mstar}{\text{M}_{\star}}
\newcommand{\Rval}{\text{R}_{\rm valley}}

\newcommand{\Zenv}{\rm Z_{\text{env}}}
\newcommand{\Mp}{\rm M_{\text{P}}}
\newcommand{\MHHe}{\rm M_{\text{HHe}}}
\newcommand{\Mw}{\rm M_{\text{H$_2$O}}}
\newcommand{\fhhe}{f_{\text{HHe}}}
\newcommand{\fw}{f_{\text{H$_2$O}}}
\newcommand{\Rp}{\rm R_{\text{P}}}
\newcommand{\Teq}{\rm T_{\text{eq}}}

\begin{document} 
 \title{A fading radius valley towards M-dwarfs,\\ a persistent density valley across stellar types\thanks{The data to generate the figures is available at ZONODO DOI: 10.5281/zenodo.10719523}}
 \titlerunning{Radius valley with stellar mass}
\authorrunning{Venturini et al.}

\author{J. Venturini\orcidlink{0000-0001-9527-2903}\inst{1},
M. P. Ronco\orcidlink{0000-0003-1385-0373}\inst{2,4}, 
O. M. Guilera \orcidlink{0000-0001-8577-9532}\inst{2,4},
J. Haldemann\orcidlink{0000-0003-1231-2389}\inst{3},
C. Mordasini\inst{3},
M. Miller Bertolami \orcidlink{0000-0001-8031-1957} \inst{2}
}


\offprints{J. Venturini}
\institute{Department of Astronomy, University of Geneva, Chemin Pegasi 51, 1290 Versoix, Switzerland, 
\and
Instituto de Astrof\'{\i}sica de La Plata, CCT La Plata-CONICET-UNLP, Paseo del Bosque S/N (1900), La Plata, Argentina.
\and
{Department of Space Research \& Planetary Sciences, University of Bern, Gesellschaftsstrasse 6, CH-3012 Bern, Switzerland}
\and
N\'ucleo Milenio de Formaci\'on Planetaria (NPF), Chile.
\\
\email{julia.venturini@unige.ch}}

\date{}

 
\abstract{The radius valley separating super-Earths from mini-Neptunes is a fundamental benchmark for theories of planet formation and evolution. Observations show that the location of the radius valley decreases with decreasing stellar mass and with increasing orbital period. Here, we build from our previous pebble-based formation model, which, combined with photoevaporation after disc dispersal, unveiled the radius valley as a separator between rocky- and water-worlds. In this study, we expand our models for a range of stellar masses spanning from 0.1 to 1.5 $\Msun$. 
We find that the location of the radius valley is well described by a power-law in stellar mass as $\Rval = 1.8197 \, \Mstar^{\!0.14({+0.02}/{-0.01})}$,
which is in excellent agreement with observations. We also find very good agreement with the dependence of the radius valley on orbital period, both for FGK- and M-dwarfs. Additionally, we note that the radius valley gets filled towards low stellar masses, particularly at 0.1-0.4 $\Msun$, yielding a rather flat slope in $\Rval - \rm P_{\rm orb}$. This is the result of orbital migration occurring at lower planet mass for less massive stars, which allows for low-mass water-worlds to reach the inner regions of the system, blurring the separation in mass (and size) between rocky- and water-worlds. Furthermore, we find that for planetary equilibrium temperatures above 400 K, the water in the volatile layer exists fully in the form of steam, puffing the planet radius up compared to condensed-water worlds. This produces an increase in planet radii of $\sim 30\%$ at 1~\ME\, and of $\sim 15\%$ at 5 \ME\, compared to condensed-water-worlds. As with Sun-like stars, we find that pebble accretion leaves its imprint on the overall exoplanet population as a depletion of planets with intermediate compositions (i.e, water mass fractions of $\sim0-20\%$), carving a valley in planet density for all spectral types (abridged).}  


   \maketitle
%

\section{Introduction}
Exoplanets with sizes between Earth and Neptune are the most abundant type known until now \citep[e.g.][]{Petigura2022}. The size distribution of these objects is bimodal , with a ``radius valley" at $\sim1.5$-2 \RE separating the smaller super-Earths from the larger mini- (or sub-) Neptunes \citep{Fulton17, Fulton18, VanEylen18, Martinez19}.
The exact location of the radius valley depends on the stellar mass and planets' orbital period, with different data analysis yielding different slopes in the $\rm log(\Rp)-\rm log(\Mstar)$ and $\rm log (\Rp)- \rm log (P_{\rm orb})$ planes \citep[with $\Rp,P_{\rm orb}$ and $\Mstar$ being the planet radius, orbital period and stellar mass, respectively;][]{Fulton17, Fulton18, Berger20, Petigura2022, Luque22, Ho23, Bonfanti23}.

Understanding the origin of the radius valley and its dependence on orbital and physical parameters has become a crucial endeavour in modern exoplanetology \citep[e.g.][]{Owen17,Ginzburg18,Gupta19, Venturini20Letter, Lee21}.
The first mechanisms to account for the existence of the radius valley were the pure evolution models known as photoevaporation \citep[e.g][]{Lopez13, Owen17, JinMord18} and core-powered mass-loss \citep[]{Ginzburg18, Gupta19}. In the view of these models, super-Earths and mini-Neptunes originate from the same single-composition population of rocky cores with H/He atmospheres. The distinction between super-Earths and mini-Neptunes arises because some planets lose their atmospheres completely during the evolution, while others retain $\sim1\%$ by mass of H/He. Both models find slopes in the radius-period and radius-stellar mass plane in agreement with observations \citep{Gupta19, Rogers21}. 
However, these pure evolution models are in contradiction with planet formation theory, which predicts most mini-Neptunes to have formed beyond the water iceline and to be therefore water-rich \citep{Alibert13,Raymond18, Bitsch18, Izidoro19, Brugger20, Venturini20Letter}. 
In \citet[][hereafter V20]{Venturini20Letter}, we performed pebble-based planet formation simulations for Sun-like stars, including a self-consistent treatment of pebble growth. We found that because of the different physical properties of icy vs. rocky pebbles, icy cores are typically born bigger than rocky ones. In terms of total planet radius, this bi-modality is hidden at the time of disc dissipation, because most planets accrete gas that puffs the planetary radii up. The separation between rocky- and water-worlds becomes clear once atmospheric mass-loss sets in, stripping the atmospheres of the small rocky and icy planets, and unveiling the primordial radius valley separating the two planetary types, at $\sim$1.5-2 \RE \citep{Venturini20Letter}. In this work, we adapt and apply our planet formation and evolution model to a wide range of stellar masses, from 0.1 to 1.5 $\Msun$ (i.e., stellar spectral types spanning from M-dwarfs to A-type). The upper limit on stellar mass is motivated by the earliest-stellar-type exoplanets discovered by Kepler \citep[e.g.][]{Berger20}. On the theoretical side, we note that no population study has addressed the origin of planets orbiting super-solar stellar-masses until now. This is important to address given the new TESS data on exoplanets around F- and A-stars \citep{Psaridi22}.

Regarding M-planets, the motivation to study them is based on the increasing number of facilities targeting such objects, such as TESS \citep{TESS_paper}, CHEOPS \citep{CHEOPS_paper} and the newly-commissioned NIRPS \citep{NIRPS_ref}. In particular, regarding the radius valley, several studies are reporting masses and radii of super-Earths/mini-Neptunes \citep[e.g.][]{Demory20, VanEylen21, Luque22, Bonfanti23}, and the mere existence of the radius valley for M-dwarfs remains controversial \citep[][]{Cloutier20,Luque22, Bonfanti23}. 
From a theoretical perspective, \citet{Alibert17}, \citet{Miguel2020} and \citet{Burn21} showed that low-mass water-rich planets should be common at close-in orbits around M-dwarfs. 

In this work, we conduct a large planet formation and evolution parameter study, focusing on the population of low- and intermediate-mass planets. The aim is to describe, theoretically, the behaviour of the radius valley for different stellar types. Our main results are highlighted in the brief main text, and several in-depth analysis are supplemented in the Appendices.

\section{Methods}\label{sec_methods}
We compute the formation and evolution of planets around stars of mass 0.1, 0.4, 0.7, 1.0, 1.3 and 1.5 $\Msun$.
In our model \citep[][]{Venturini20Letter, Venturini20SE}, planets grow from a lunar-mass embryo by pebble and gas accretion, while embedded in a gaseous disc that evolves by viscous accretion and X-ray photoevaporation from the central star. The discs are adapted to different stellar masses following \citet{Burn21} (details in Appendix \ref{App_IniCond}).
The evolution of the pebbles is calculated considering pebble coagulation, fragmentation, drift, diffusion, and the ice sublimation at the water iceline \citep[model of][]{Birnstiel11, Drazkowska16}.
The planets can accrete either rocky or icy pebbles, depending on its position with respect to the water iceline (which evolves in time, moving inwards as the disc cools down). Core growth halts when pebble isolation mass is reached \citep{Lambrechts14}. As in V20, we adopt a fragmentation threshold velocity of icy and silicate pebbles as v$_{\rm th}=10$ m/s and v$_{\rm th}=1$ m/s, respectively \citep[e.g.][]{Gundlach15}. We note that, based on experimental work, these values are still quite uncertain (see detailed discussion in App.\ref{App_frag_pebbles}). Our results hold as long as the fragmentation threshold velocity of icy pebbles is in the range of 5-10 m/s. More experimental work and sensitivity studies are required to better constrain these fragmentation velocities.
For the disc viscosity, we consider $\alpha$ either as $10^{-3}$ or $10^{-4}$. The details about the choice of the initial conditions are given in App. \ref{App_IniCond}.
The planets also migrate along the disc, either via the type I-or type-II regime. The former includes the Lindblad, corotation and thermal torques \citep{Guilera19}. 

As explained in V20, since our code can handle at the moment the formation of only one embryo per disc, and given the fast timescales of pebble accretion and type-I migration, most of our formed planets get stranded near the disc inner edge, which is set at an orbital period of 11.55 days for all the considered discs (stemming from the choice of disc inner edge at 0.1 au for solar-type stars, see App. \ref{App_IniCond}). This means that most of the results reported throughout this work refer to planets with orbital period of $\sim$11-18 days.

Because of the limitation of considering only one embryo per disc, the possibility of  atmospheric mass loss due to giant impacts is in principle hindered in our model. Nevertheless, as in V20, we estimate this rate in a simplified way, following \citet{Inamdar15} and \citet{Ronco17}, by considering a potential impact between the simulated planet and another one less massive formed in a different simulation, but under the same disc initial conditions (see App.\ref{App_IniCond}). We refer to this batch of simulations as the "collision case" (while the standard simulations without the estimation of atmospheric mass-loss by collisions is referred as "nominal case"). The details of this implementation and its limitations can be found in App.\ref{App_collisions}. Once the disc dissipates, the cooling of the planets (including the cooling of the core) is calculated during 2 Gyr with photoevaporation following \citet{Mordasini20}. We refer to this phase as "evolution", while with "formation" we refer to the disc-embedded phase when planetary accretion takes place. During evolution, we consider the planets to consist of a rocky core of Earth-like composition, surrounded by an envelope of volatiles which can contain mixtures of H/He with water (with all the components and amount stemming from the formation process). The atmospheric mass-loss occurs for all the components of the volatile layer. Further details of the evolution model are described in App.\ref{App_evol}. Finally, we compute the planetary radii after 2 Gyr of evolution with state-of-the-art interior structure models (details in App.\ref{App_biceps}). The evolution and structure calculations are performed for all the planets that ended with orbital periods below 100 days after formation, and these are the cases analysed and presented throughout this study.

\section{Results}
\subsection{Mass-Radius diagrams: general trends}\label{sect_MR}   
   \begin{figure*}[ht]
   \begin{center}
   \includegraphics[width=1\linewidth]{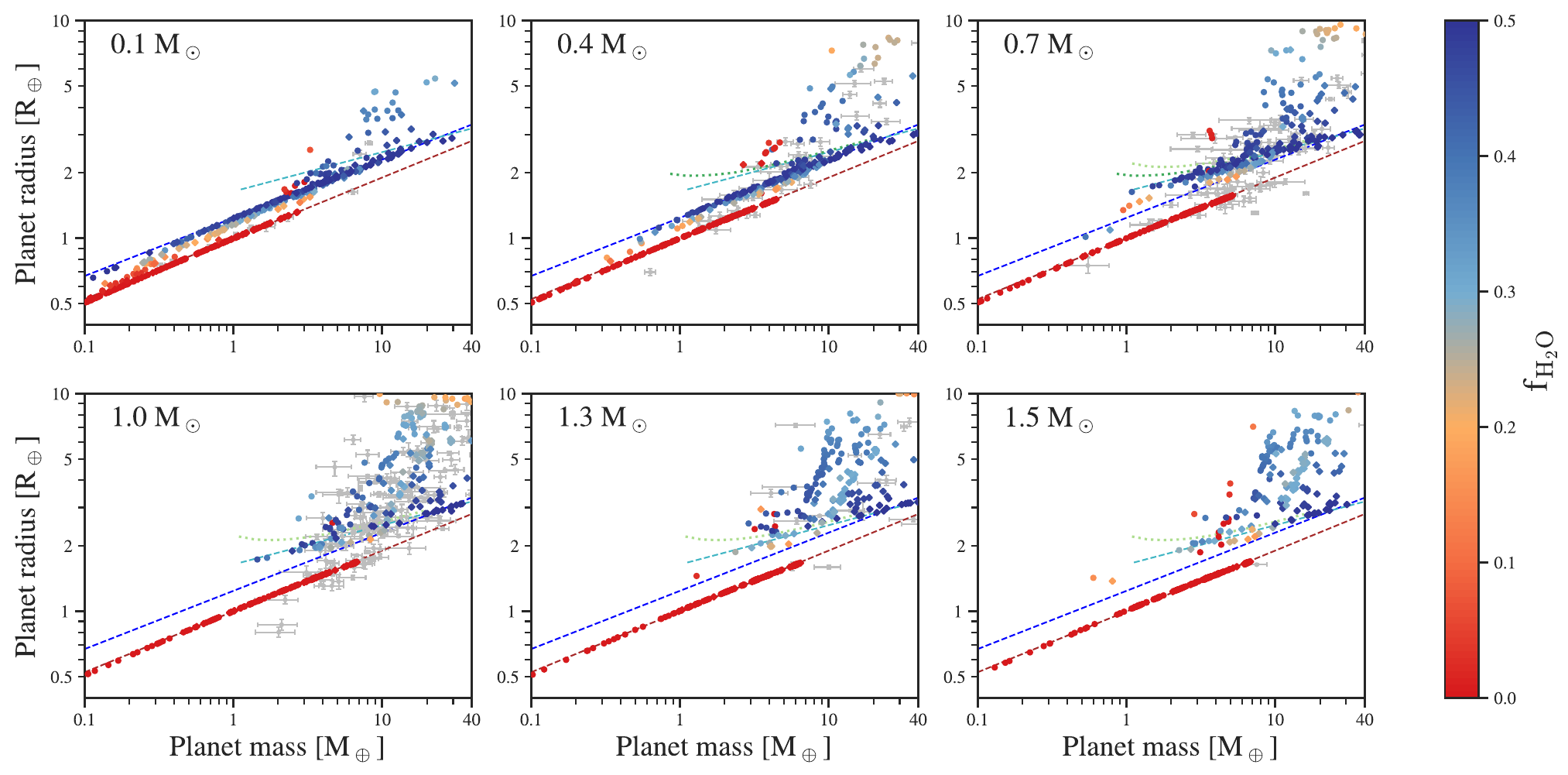}
   \caption{Mass-Radius (M-R) diagram for the different stellar masses. The colour-bar represents the total water mass fraction in the planets. Dots correspond to the nominal output, diamonds to collisions. The brown- and blue-dashed lines correspond to a theoretical composition of Earth-like and 50\% water-50\% Earth-like, as defined in Ap.\ref{app_MRfits}, with dark-blue indicating condensed-water and light-blue, steam-worlds. The green-dotted curves correspond to 50\% steam-50\% Earth-like from \citet{Aguichine21} ($\Teq=400$ K for dark-green and $\Teq=600$ K for light-green). Grey dots are real exoplanets mass and radius measurements better than 75 and 20\%, respectively (data taken from the NASA Exoplanet Archive on 08.09.23. The real exoplanets in each panel correspond to the stellar masses of 0-0.25 $\Msun$ for $\Mstar=0.1 \,\Msun$, 0.25-0.55 $\Msun$ for $\Mstar=0.4 \,\Msun$, 0.55-0.85 $\Msun$ for $\Mstar=0.7 \,\Msun$, 0.85-1.15 $\Msun$ for $\Mstar=1.0 \,\Msun$,1.15 -1.4 $\Msun$ for $\Mstar=1.3 \,\Msun$, and 1.4-1.7 $\Msun$ for $\Mstar=1.5 \,\Msun$.}
              \label{Fig_MRfice}%
     \end{center}
    \end{figure*}
We analyse the output of our formation-evolution simulations at time 2 Gyr. Figure \ref{Fig_MRfice} shows the M-R diagram at that time, for the different stellar masses, colour-coded with the planet's water mass fraction (total water mass divided by the total planet mass). The dotted lines represent the M-R relations for an Earth-like composition (brown), and for planets composed of 50\% water and 50\% rocks by mass, with the water either in condensed form (blue) or in the form of steam (light-blue). Hereafter, we refer to these compositional curves as the "condensed water-line" and the "steam-line", respectively. The derivation of these M-R relations is detailed in Sect.\ref{app_MRfits}.

A general trend that we find for all panels, is a division given by the rocky and condensed water lines. Bare rocky cores cluster around the Earth-like composition (as they should by hypothesis) and wet planets tend to be either on the condensed water-line or above. This diagonal band between the rocky and condensed water lines remains planet-depleted for all stellar masses. (Although some planets with intermediate compositions exist in between those lines, particularly for M-dwarfs, see Fig.\ref{Fig_MRfice}).  Planets with radii larger than the one given by the condensed water-line tend to have substantial H/He, particularly when moving towards larger planet mass. This can be appreciated as a fading of the blue colour towards large planetary radii (the mass fraction of water decreases as the H/He mass fraction increases). The H/He mass of each synthetic planets is also depicted in Fig. \ref{Fig_MRfhhe}.  
Another clear trend of Fig.\ref{Fig_MRfice} is that for $\Mstar\ge0.7\,\Msun$, a large number of water-worlds cluster around the steam-line (particularly for $\Mstar=0.7\,\Msun$). Indeed, for those cases the temperatures are high enough for water to be present in the form of steam throughout the atmosphere. This is demonstrated in Sect. \ref{app_steam}. Other aspect analysed in detail in Sect.\ref{App_red-outliers} is the presence of "red" outliers among "blue" planets. These are planets that actually formed beyond the iceline but lost most of their water during evolution.

We also note from Fig.\ref{Fig_MRfice} that the lower the stellar mass, the larger the overlap in mass and radius of dry and wet planets. As the stellar mass increases, the minimum mass of wet planets increases as well, taking values of $M_{\rm min} \simeq $ 0.6, 1.4, 2.8, 4.5, 5.4 and 7.3 \ME \, for $\Mstar= 0.1, 0.4, 0.7, 1.0, 1.3, 1.5 \, \Msun$, respectively\footnote{$M_{\rm min}$ is taken as the 5\% percentile of the distribution of wet planets, to avoid a few outliers.}. This is the imprint of orbital migration. Indeed, the lower the stellar mass, the lower the threshold mass to undergo type-I migration when embedded in a gaseous disc \citep{Paardekooper2011}. The same effect was reported in \citet{Burn21}. This produces a large overlap in mass and radii between dry and wet planets for $\Mstar = 0.1 \, \Msun$, filling the "radius valley" as noted in the histograms of Fig. \ref{Fig_Ap2} for $\Mstar \lesssim 0.4 \, \Msun$. The radius valley starts to clearly emerge for $\Mstar \gtrsim 0.7 \, \Msun$, where water-worlds have typical masses above 2-3\ME. 

\subsection{The valley in the radius-stellar mass plane}\label{sec_rvalley_mstar} 
\begin{figure}[h]
    \centering
    \includegraphics[width= 1.0\columnwidth]{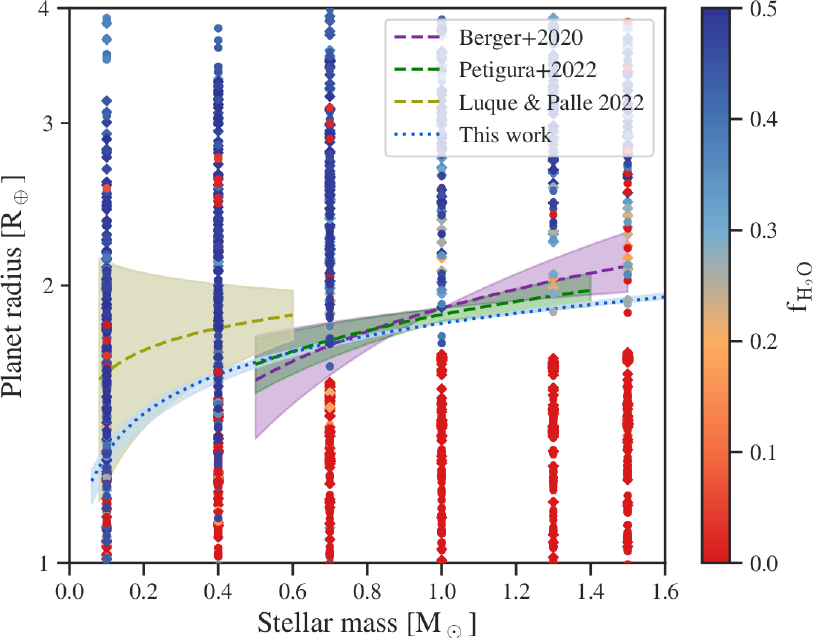}
    \caption{Radius valley fit as a function of the stellar mass (blue line with errors in light-blue) for all our synthetic planets (colour-coded as a function of the water mass fraction). Dots correspond to the nominal output, diamonds to collisions. The green dashed line with the light-green shaded area show the best fit found by \citet{Petigura2022}, the violet dashed line and the lilac shaded area represent the results found by \citet{Berger20}, and the mustard dashed line and the light-mustard shaded area represents the best fit by \citet{Luque22}. Our best fit was found using \url{gapfit}\citep{Loyd2020}.}
    \label{fig:rp_vs_mp}
\end{figure}
Different analysis of exoplanets' data have shown that the radius valley follows a power-law dependence with stellar mass, i.e, \begin{equation}
 \Rval= \text{R}_{\rm v,0} \left(\dfrac{\text{M}_{\star}}{\text{M}_{\odot}}\right)^{m}.  
 \label{eq:obs_gap_position_Rp_vs_Mstar}
\end{equation}
For KGF stars from the \textit{Kepler} data, \citet{Berger20} found $d {\rm log}{\rm \Rval}/d {\rm log}\Mstar = 0.26^{+0.21}_{-0.16}$, and \citet{Petigura2022}, $\text{R}_{\rm v,0} =1.86^{+0.03}_{-0.03}~\text{R}_{\oplus}$, $m= 0.18^{+0.08}_{-0.07}$.  
For M-dwarfs, \citet{Luque22} found $d {\rm log}{\rm \Rval}/d {\rm log}\Mstar = 0.08\pm0.12$.

In Fig.~\ref{fig:rp_vs_mp}, we plot the planet radius as a function of the stellar mass for our synthetic planets, colour-coded by $\fw$, as in Fig.\ref{Fig_MRfice}. 
The blue-dotted line represents our best fit of the valley adopting the power-law dependence of Eq.\ref{eq:obs_gap_position_Rp_vs_Mstar}. The location of the valley was found using the open tool \url{gapfit} \citep{Loyd2020}. 

\begin{figure}[h!]
    \centering
        \includegraphics[width=1.0\linewidth] {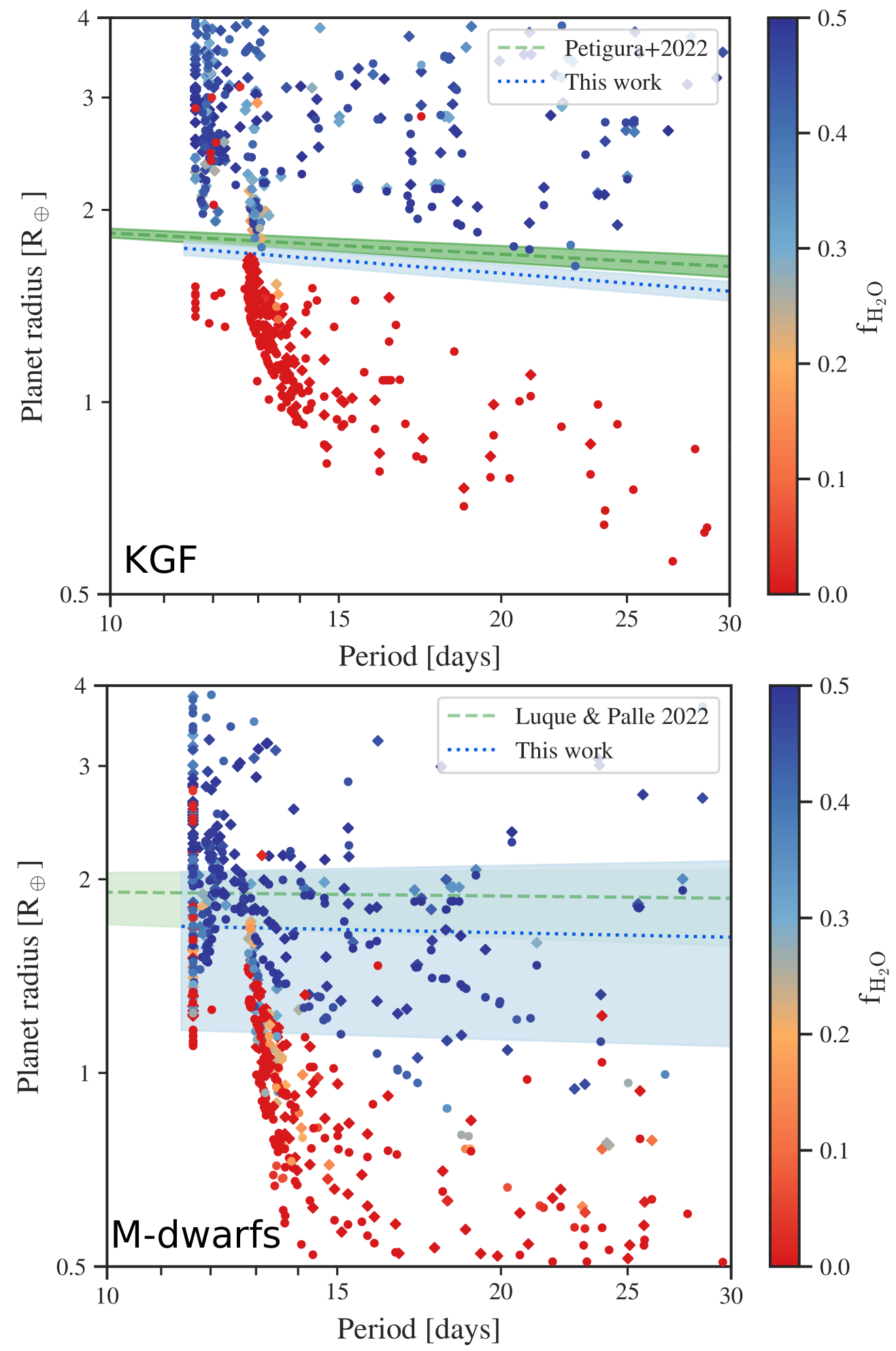}
    \caption{Radius valley fit as a function of orbital period (blue line with errors in light-blue). The colour-bar indicates total water mass fraction. \textit{Top panel:} planets around KGF-type stars. The green  line and shade show the best fit found by \citet{Petigura2022} for these stellar types. \textit{Bottom panel:} planets around M-dwarfs. The green line and shade show the best fit found by \citet{Luque22}. The particular clustering in orbital period in both panels is due to model limitations ( App.\ref{app_model_limitations}).}
    \label{fig:Rp_Porb}
\end{figure}

By adopting  $\text{R}_{\text{V}_0,\text{guess}}=1.82$ and $m_{\text{guess}}=0.09$ in \url{gapfit}, our best fit yields $\text{R}_{\rm v,0}=1.8197$ and $m=0.14^{+0.02}_{-0.01}$. We note that we are not plotting the $1\sigma$ error computed with \url{gapfit}, but the best fit with the corresponding errors of the constants $\text{R}_{\rm v,0}$ and $m$. 
We also over-plot fits from previous studies with their corresponding errors in Fig.~\ref{fig:rp_vs_mp} (see figure caption).
Our theoretical results are in excellent agreement with the observational results, for all the stellar masses ranging from 0.1 to 1.5 $\Msun$.  
Fig.~\ref{fig:rp_vs_mp} also shows that for $\Mstar \lesssim 0.4~\Msun$, the radius valley fades (or, in other words, it gets filled), because of the increasing overlap between rocky- and water-worlds (see Fig.\ref{Fig_MRfice}). A vanishing radius valley for M-dwarfs is observationally supported by the findings of \citet{Luque22} and \citet{Ho2024}. 
Another remarkable aspect of Fig.~\ref{fig:rp_vs_mp} is that for $\Mstar \gtrsim 0.7~\Msun$, the radius valley is really empty, particularly for $\Mstar\geq 1.3~\Msun$. This is in agreement with the findings of  \citet{VanEylen21} and \citet{Ho23}.

\subsection{The valley in the radius-orbital period plane}\label{sec_rvalley_Porb} 

We additionally analysed the location of the valley as a function of orbital period using \url{gapfit} for KGF-stars and M-dwarfs. The result of the best fit for KGF-stars is $\Rval (\text{R}_{\oplus})=2.5704\,\rm{P_{orb}^{{-0.16 \pm0.01}}}$ and is shown in Fig.\ref{fig:Rp_Porb} together with the fit derived by \citet{Petigura2022}, $\Rval (\text{R}_{\oplus}) =2.37^{\pm0.04}\,\rm{P_{orb}}^{-0.11 \pm0.02}$. We note that including A-type stars ($\Mstar=1.5 \,\Msun$) does not alter the fit. 
Despite the limitations of our model (too many planets concentrated at $\rm P_{orb} \approx 11-15$ days, see Sects.\ref{sec_methods} and \ref{app_model_limitations}), we find a remarkable good agreement with observations, with a negative slope carved by photoevaporation. For M-dwarfs ($\Mstar=0.1 \,\rm{and} \, 0.4 \, \Msun$), we find a slope of $-0.04^{+0.08}_{-0.03}$, i.e. a flat slope, in line with \cite{Luque22} --who found a slope of $-0.02\pm0.05$--, and also in agreement (within errors) with \citet{Bonfanti23}, who derived a slope of $-0.065^{+0.024}_{-0.013}$. However, we caution that in this case the package \url{gapfit} cannot easily find a gap, with different initial guesses yielding different slopes. This is not a problem of lack of points --with other pairs of stellar bins this problem does not arise--, which reinforces the conclusion that the valley gets indeed filled when moving towards M-dwarfs.  

\section{Discussion: the compositional valley}\label{sect_disc}
\begin{figure*}[h]
    \centering
    \includegraphics[width=\linewidth]{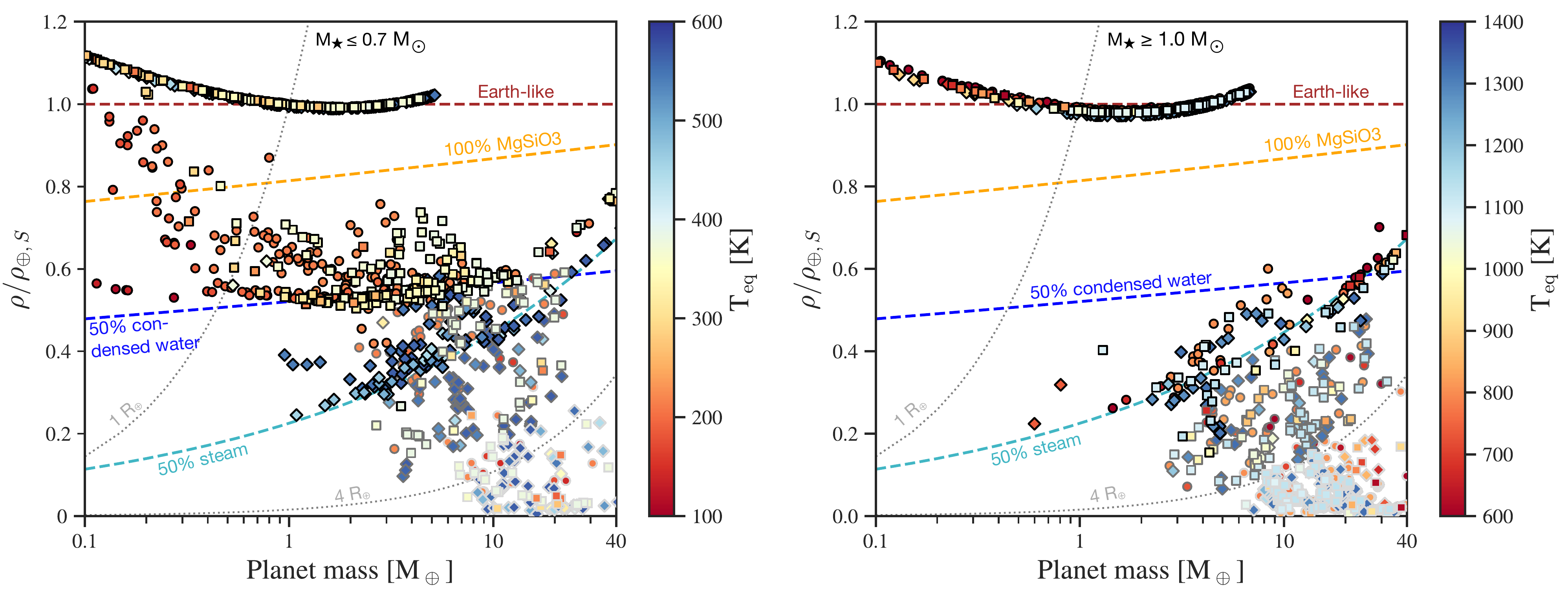}
    \caption{Density-Mass diagram (density normalised by Earth-like composition) as a function of equilibrium temperature (colour-bar). \textit{Left panel}: planets orbiting stars of 0.1 $\Msun$ (circles), 0.4 $\Msun$ (squares) and 0.7 $\Msun$ (diamonds). \textit{Right panel}: planets orbiting stars of 1.0 $\Msun$ (circles), 1.3 $\Msun$ (squares) and 1.5 $\Msun$ (diamonds). Note that the colour-bars span different ranges in the two panels. For both panels, black-bordered symbols indicate $\fhhe<1\%$, dark-grey $1\leq\fhhe\leq10\%$ and light-grey $\fhhe>10\%$. The dashed-lines show the compositional curves of Earth-like (brown), pure-silicate (orange), condensed water (blue), and steam worlds (light-blue). The grey-dotted lines correspond to radii of 1 and 4 \RE.}     \label{fig_discussion}
\end{figure*}
We computed the growth and post-formation evolution of planets orbiting starts from 0.1 to 1.5 $\Msun$. Several model limitations exist, which we discuss in detail in App.\ref{app_model_limitations}. 
Overall, our findings reinforce our conclusion of V20 that the radius valley is a separator in planetary composition between bare rocky cores whose atmospheres were stripped off by photoevaporation, and water-worlds that formed beyond the iceline and retained some or none of their primordial H/He. However, the radius valley fades with decreasing stellar mass (Figs.\ref{Fig_MRfice},~\ref{fig:rp_vs_mp},~\ref{fig:Rp_Porb}). This is a consequence of orbital migration, which happens for less massive planets at lower stellar masses \citep{Paardekooper2011, Burn21}. Hence, for low mass-stars, originally small icy planets formed beyond the iceline can reach the inner regions of the system, rendering a larger overlap in mass and radius between planets formed inside and outside the iceline for M-dwarfs.

Nevertheless, the paucity of planets with intermediate values of water mass fraction ($0<\fw\lesssim 0.2$) persists for all stellar masses, yielding an depleted ($\Mstar\le 0.4\, \Msun$) or empty ($\Mstar\ge 0.7\, \Msun$) diagonal band in the M-R diagram of Fig.\ref{Fig_MRfice} between the rocky- and condensed water- lines. This is due to pebble accretion, which proceeds quickly beyond the iceline, allowing the planets to typically reach the pebble isolation mass before crossing it. This means that solid accretion tends to stops before the icy planets have the possibility to accrete dry pebbles and mix their bulk composition (particularly true for $\Mstar\gtrsim1.0\,\Msun$). If this "diagonal valley" were not confirmed by future observations, then planetesimal-based- \citep[e.g.][]{Brugger20} or hybrid pebble-planetesimal models \citep{A18, Guilera20} would be favoured. 
This "diagonal valley" in the M-R diagram appears more clearly when plotting, instead of radius, mean density \citep[normalised by the density of the Earth-like composition, as in][see as well App.\ref{app_density}]{Luque22}.
We show this in Fig.\ref{fig_discussion}, including the compositional lines of Earth-like, pure-silicate composition, condensed-water and steam-worlds. The colour-bars indicate the planetary equilibrium temperature, and the different symbols the different stellar masses (0.1-0.7 $\Msun$ on the left panel, 1.0-1.5 $\Msun$ on the right). 
A clear pattern emerging from Fig.\ref{fig_discussion} is a depletion (void) of planets with $\rho \approx 0.6-0.9 \, \rho_{\oplus,S}$ around stars of 0.1-0.7 $\Msun$ (1.0-1.5 $\Msun$). Our model assumed an Earth-like composition for all the rocky components. In reality, different stellar abundances would yield rocky planets with different iron fractions, implying that rocky planets could be expected down to the pure silicate line (in orange) of Fig.\ref{fig_discussion}. This means that in reality the density valley would be more blurred, but should still exist at $\rho \approx 0.6-0.8 \,\rho_{\oplus,S}$ as a consequence of pebble accretion. This is overall in line with the conclusions of \citet{Luque22}, who found a valley at $\rho \approx 0.65 \,\rho_{\oplus,S}$ for M-planets. 

We note as well that water-worlds with \textit{condensed} water exist mainly for $\Mstar= 0.1 \, {\rm and} \, 0.4\, \Msun$, where the equilibrium temperatures are below $\sim 400$~K (Fig.\ref{fig_discussion}, bottom left, and Fig.\ref{fig_PTwaterWorlds}). For 0.7 $\Msun$, water-worlds are very common, but water is present as steam along all the planetary envelope. Thus, for planets with $\Teq \gtrsim 400$~K, we predict that water worlds should cluster around the line of 50\% rock-50\% steam by mass (at approx. 0.2-0.4 $\rho_{\oplus,S}$), as long as planetary cores build predominately from pebbles. This steam-line has planetary radii larger than the condensed water line by $30\%-15\%$ for planetary masses of $1-5 \,\Mearth$, following the tendency reported by \citet{Turbet19} and \citet{Aguichine21}. This means that the existence of steam-worlds helps to carve a deeper radius valley for $\Teq \gtrsim 400$~K \citep[also reported by][for planets around Sun-like stars, with $\Teq \gtrsim600$ K]{Burn2024}. In this regard, our results do not match entirely the ones of \citet{Luque22}, who report water worlds of $\Teq > 400$~K falling on the condensed-water line. A possible explanation might be steam atmospheres with a reduced water mass fraction (see discussion in App.\ref{App_real_vs_synt}). Increasing the sample of M-planets will be crucial to test our model predictions.

Finally, Figs. \ref{Fig_MRfice} and \ref{fig_discussion} also illustrate how mini-Neptunes span from being predominantly water-condensed worlds for $\Mstar=0.1 \,\rm{and}\, 0.4  \,\Msun$, to steam-worlds for $\Mstar= 0.7 \,\Msun$, to steam-H/He planets for $\Mstar \gtrsim 1.0 \, \Msun$, with a H/He mass fraction below 10\%. This prediction holds within the ranges in equilibrium temperature considered in this work (see Fig.\ref{fig_discussion}). The case of $\Mstar = 1.5 \,\Msun$ is somehow the exception, where several mini-Neptunes are dry due to the high loss of water during evolution (see Fig.\ref{Fig_MRfice} and Sect.\ref{App_red-outliers}).

\section{Conclusions}
We studied the combined formation and evolution of planets around single stars in the mass range $0.1\leq\Mstar\leq1.5 ~\Msun$, with the goal of characterising the radius valley for different stellar types. Overall, we find that the tendency found in V20 to have larger water-worlds compared to smaller rocky planets, persists for all stellar ranges, but the radius valley separating them fades towards M-dwarfs \citep[in agreement with][]{Luque22, Bonfanti23}. This is a consequence of orbital migration, which allows small icy planets to reach the disc inner regions for low-mass stars, producing a larger overlap in mass and radius between the rocky and icy populations. 

Despite of the "filling" of the radius valley towards low stellar masses, we find that when considering the full range of stellar masses, its dependence on stellar mass is in excellent agreement with observations, following $d \text{log}\Rval/d\Mstar = 0.14^{+0.02}_{-0.01}$.
We find as well a good agreement with observations regarding the location of the radius valley with orbital period (see Sect.\ref{sec_rvalley_Porb}), with a negative slope carved by photoevaporation for KGF-stars, and a flat slope for M-dwarfs.

Our end-to-end simulations show that the super-Earths are bare rocky cores that emerge as evaporated worlds around all stars. The mini-Neptunes, on the other hand, are in their vast majority water-worlds that lost all or part of their primordial H/He due to photoevaporation.
In addition, we confirm that the phase of water is key in shaping the radii of the exoplanet population. If the outer atmospheric temperatures are low enough for water to condense ($\Teq \lesssim 400$~K), then a water-world will transition from a steam- to an icy/liquid-world, reducing its size by $\sim 15\%$ at a planet mass of 5 \ME. 

Finally, because within our pebble-based formation and evolution model the radius valley emerges as a divider in planetary composition, we note that it is more visible in terms of mean density, where a clear valley occurs at normalised mean density of $\rho/\rho_{\oplus,S}\sim 0.6-0.8$ across all stellar types. 

\begin{acknowledgements}
We thank the anonymous referee for insightful criticism. JV acknowledges support from the Swiss National Science Foundation (SNSF) under grant PZ00P2$\_$208945. JH and CM acknowledge support from the Swiss National Science Foundation (SNSF) under grant 200021\_204847 ``PlanetsInTime''. MPR is partially supported by PICT-2021-I-INVI-00161 from ANPCyT, Argentina. MPR, OMG and MMB are partially supported by PIP-2971 from CONICET (Argentina) and by PICT 2020-03316 from Agencia I+D+i (Argentina). MPR  and OMG acknowledge support by ANID, -- Millennium Science Initiative Program -- NCN19\_171. OMG gratefully acknowledges the invitation and financial support from ISSI Bern in early 2024 through the Visiting Scientist Program. 
\textit{Software.} For this publication the following software packages have been used: 
\href{https://matplotlib.org/}{Python-matplotlib} by \citet{Hunter_2007}, \href{https://seaborn.pydata.org/}{Python-seaborn} by \citet{waskom2020seaborn},
\href{https://numpy.org/}{Python-numpy},
\href{https://pandas.pydata.org/}{Python-pandas}.

\end{acknowledgements}

\bibliographystyle{aa}
\bibliography{lit_2021}

\begin{appendix}
\section{Methodology}
\subsection{Formation model: initial conditions}\label{App_IniCond}
We apply the global planet formation model {\scriptsize PLANETALP} \citep{Ronco17,Guilera17, Guilera19} with the stellar masses 0.1, 0.4, 0.7, 1.0, 1.3, and $1.5~\text{M}_{\odot}$. As in V20, we use the 12 protoplanetary disc profiles from \citet{Andrews10} with 2 uniform viscosity parameters ($\alpha=10^{-4}, 10^{-3}$), where the initial gas surface density is given by:
\begin{equation}
  \Sigma_{\text{g}} = \Sigma_{\text{g,0}} \left( \frac{r}{r_\text{c}} \right)^{-\gamma} e^{-(r/r_\text{c})^{2-\gamma}},
  \label{eq1_app_a1}
\end{equation}  
where $\Sigma_{\text{g,0}}$ is a normalisation parameter determined by the disc initial mass ($\text{M}_{\text{d,0}}$), $\gamma$ is the exponent that represents the surface density gradient, and $r_\text{c}$ the characteristic or cut-off radius of the disc. Following \citet{Burn21}, we scale the initial disc mass and characteristic radii as:
\begin{eqnarray}
\text{M}_{\text{d,0}} &=& \left( \dfrac{\text{M}_{\star}}{\text{M}_{\odot}} \right) \, \text{M}_{\text{d,0}}^{\rm sun},\\
r_{\text{c}} &=&  \left( \dfrac{\text{M}_{\star}}{\text{M}_{\odot}} \right)^{1/1.6} r_{\text{c}}^{\rm sun}, 
\end{eqnarray}
where $\text{M}_{\text{d,0}}^{\rm sun}$ and $r_{\text{c}}^{\rm sun}$ stem from \citet{Andrews10} \citep[displayed in Tab. B.1 of][]{Venturini20Letter}.
In \citet{Venturini20Letter} we considered only solar mass stars and we set the inner edge of the disc at 0.1~au, based on hydrodinamical simulations \citep{Flock19} and on the mean orbital period of the innermost planet of planetary systems \citep{Mulders18}. Also as in \citet{Burn21}, we scale the disc inner edge as:
\begin{eqnarray}
\text{r}_{\text{int}} &=&  \left( \dfrac{\text{M}_{\star}}{\text{M}_{\odot}} \right)^{1/3} \,  0.1~\text{au}.
\end{eqnarray}
This scaling is based on identifying the location of $\text{r}_{\text{int}}$ as the point where the Keplerian orbital period matches the stellar rotation period \citep[assumed to be the same for all stars as in][]{Burn21}. However, we note that the location of the disc inner edge and its scaling with stellar mass is very uncertain.

Given the initial gas surface density by Eq.~(\ref{eq1_app_a1}), the initial dust surface density is defined as 
\begin{equation}
    \Sigma_{\text{d}}= \eta_{\text{ice}} Z_{0} \Sigma_{\text{g}}
    \label{eq:eq8-sec2-1}
\end{equation}
where $\eta_{\text{ice}}$ represents the sublimation/condensation of water-ice and adopts values of $\eta_{\text{ice}}$=1/2 inside the iceline and  $\eta_{\text{ice}}$=1 outside of it \citep{Lodders09}. $Z_{0}$ is the initial dust-to-gas ratio, which adopts the values of 0.0068, 0.0099, 0.0144, 0.0210 and 0.0305, to span the known ranges of stellar metallicities (as assumed in V20). 

Regarding the thermodynamical quantities of the discs around stars with different masses, we use the corresponding effective temperatures $\text{T}_{\text{eff}}$ and stellar radii $\text{R}_{\star}$ from \citet{Baraffe15} for the different protostars at 0.5~My to compute the disc vertical structure \citep[as in][]{Guilera17b, Guilera19}. 

As in \citet{Venturini20Letter} we launch seven embryos per disc (one at a time), 3 of them inside and 4 of them beyond the iceline. For the embryos that start within the iceline in the solar-case, the initial semimajor axis is defined with uniform log-spacing between $\text{r}_{\text{int}}$ and $r_\text{ice}-0.1$ au, and between $r_\text{ice}+0.1$ au and 16 au for the icy embryos. For the non-solar cases, the corresponding initial semimajor axis scales as $a_{\text{ini}}(\text{M}_{\star}/\text{M}_{\odot})^{1/3}$ as the disc inner edge.

\subsection{Gas accretion: calculation and assumptions}\label{App_gas_acc}
Gas accretion is computed both in the attached and detached phases as in \citet{Venturini20SE}. During the attached phase (where the planet’s envelope connects smoothly with the gaseous disc), gas accretion is calculated from solving the 1-D spherically symmetric structure equations for the envelope before the planet reaches the pebble isolation mass (following the method of \citet{Alibert19}, which uses deep neural networks trained on pre-computed structure models for sub-critical core masses). When solving the structure equations, the envelope opacity is taken from \citet{Freedman14} for the gas, and from \citet{BL94} for the dust (see discussion on the choice of opacities on App.\ref{App_opac}.) Once the pebble isolation mass is attained, solid accretion stalls and the core becomes critical. At this stage we adopt the prescription of \citet{Ikoma00}, which is valid precisely in this regime when solid accretion is halted. After reaching the pebble isolation mass, at a certain point the planet would accrete more gas than what the disc can provide due to its viscosity. Here the planet detaches from the disc and accretes at a rate given by the disc viscous accretion ($3\pi\nu\Sigma$). In the detached phase the gas accretion can be damped even further when the planet opens a gap. This is taken into account following Eqs.(36)-(39) of \citet{Tanigawa07}.

\subsection{Gas removal by giant impacts}\label{App_collisions}


Once the gas of the disc has completely dispersed, and during the first million years of evolution, giant collisions between the remaining planets/embryos may become an important mechanism to efficiently remove the planet's atmosphere \citep{Ogihara2020a, Ogihara2020b} before atmospheric mass loss due to photoevaporation takes place \citep{Izidoro2017}. 
Here, as in \citet{Venturini20Letter}, we form 7 embryos per disc but only one at a time, which implies that N-body interactions between protoplanets embedded in the gaseous disc are not modelled. However, we can simply estimate the atmospheric mass-loss due to potential collisions between the formed planet and another one less massive with final period <100 days, also formed in the same disc, but in a different simulation. For the sake of simplicity and in alignment with \citet{Ogihara2020a} findings, who observe one or two giant impacts in their N-body analysis, we restrict each planet to a single collision. We follow the procedure developed in \citet{Ronco17} where the core mass of the planet after the collision is the sum of the target's and impactor's cores, and where the final envelope mass ($M_\text{E}$) is estimated with the global atmospheric mass-loss fractions ($X_\text{loss}$) computed in \citet{Inamdar15} (their fig.5) between super-Earths and mini-Neptunes. Thus, if $M^i_\text{E}$ and $M^j_\text{E}$ are the envelope masses of the target $i$ and the impactor $j$, the envelope mass of the target after the collision is given by $M^i_\text{E} = M^i_\text{E}(1-X^i_\text{loss})+M^j_\text{E}(1-X^j_\text{loss})$, where $X^i_\text{loss}$ and $X^j_\text{loss}$ are the atmospheric mass-loss fractions of the target and the impactor, respectively. For each family of collisions (i.e, for all the isolated collisions between a planet and each of the other less massive planets in the same disc), we keep the most destructive one that removes the greatest amount of atmosphere. 

The main caveat of our simple collision model is clearly the lack of N-body simulations since we can only handle one planet per simulation. Our approach to calculating potential collisions post-gas stage with other (less massive) planets formed under the same initial disc conditions in different simulations does not guarantee their simultaneous formation in a single simulation.  On the other hand, considering only one collision per planet, particularly the most destructive one (the one that generates the maximum envelope mass-loss), could reduce the mass of the envelope between ~16\%-100\% with a mean value of ~72\%, as stated in App.D of V20. This consideration could be overestimating the envelope mass-loss rate since \citet{Ogihara2020a} showed that their mass-loss fraction ranges from a few percent to approximately 90\%, but with a typical value around 20\%. However, they also confirm that this percentage is higher if head-on collisions take place. 
On another note, \citet{Matsumoto2021}, who conducted N-body simulations to trace planet sizes during the giant impact phase with envelope stripping through impact shocks, considered empirical envelope-loss rates obtained from smoothed particle hydrodynamics simulations by \citet{Kegerreis2020}. These authors also mentioned that head-on collisions can lead to increased rates of envelope mass loss. Additionally, their findings indicate that protoplanets in inner orbits undergo a higher frequency of collisions, potentially leading them to become bare cores after giant impacts, an outcome that we find within our simulations. Thus, our model captures the overall outcomes observed in more detailed studies, despite its simplicity and reduced precision in modeling collisions.

\subsection{Evolution model}\label{App_evol}
After disc dispersal and after removing mass from giant impact (for the "collision-case"), we compute the cooling and contraction of the planets' envelopes, including the effect of mass-loss due to photoevaporation during 2 Gyr with the code {\scriptsize COMPLETO} \citep{Mordasini20}. We refer to this stage as the "evolution phase". The choice of 2 Gyr is to ensure that all the stars considered in this study are still in the main sequence. 
The code solves the 1D equations for mass, momentum, and
energy conservation. When calculating the
luminosity evolution, the gravitational and internal energy of the
core and envelope, as well as radiogenic heating are considered \citep{Linder19}. The stellar XUV stellar X-ray flux evolution is taken from the model of \citet{Mcdonald2019}.
The envelopes are assumed to be composed of H, He and H$_2$O (with the compounds uniformly mixed), and with the initial amounts stemming from the accretion process. 
The EOS of the H/He component correspond to \citet{Chabrier19} and for water we use the AQUA EOS from \citet{haldemann_2020}. The equation of state (EOS) for the iron-silicate core is the modified polytropic EOS of \citep{Seager07}, which does not yield very precise planetary radii for sub-Earth size planets, reason for which we re-computed the final radius at 2 Gyr with the more-up-to-date interior structure model of {\scriptsize BICEPS} (see next section).
 Atmospheric escape rates are taken from \citep{Kubyshkina21}. The atmospheric escape follows a dependency of $\Zenv^{-0.77}$ \citep{Owen12, Mordasini20}, where $\Zenv$ is the water mass fraction of the envelope, i.e., $\Zenv = \frac{\Mw}{\Mw + \MHHe}$. We caution that the evaporation rates of envelopes enriched in heavy elements are still  uncertain. It is imperative that future hydrodynamic escape calculations address this aspect. 

\subsection{Transit radius calculation and internal structure}\label{App_biceps}
At the end of the planetary evolution stage, we calculate for each modelled planet its corresponding transit radius. For that we use the internal structure model of {\scriptsize BICEPS} by \citet{haldemann_2023b}. {\scriptsize BICEPS} solves the internal structure equations and calculates the planetary transit radius for a large range of planetary compositions. For this work we assumed that all planets are structured in the following way: i) in the centre there is an iron core made out of pure iron, ii) surrounded by a mantle of rocky material (MgSiO$_3$), iii) the outermost layer contains the volatile elements H, He and H$_2$O. The composition of the volatile layer is given by $M_\text{HHe}$ and $M_\text{H$_2$O}$ which are provided for each planet by the evolution model. As in V20 the water is uniformly mixed with the H, and He, thus $\Zenv$ is uniform throughout the volatile layer. For planets which lost all their volatiles during the evolutionary stage or which did not accrete any volatiles, the structure model starts the integration at the mantle layer. We note that the mass ratio between iron core and rocky mantle is kept at Earth-like values of 1:2.

Depending on the planet's internal luminosity, the orbital distance to its host star and the host stars effective temperature the thermal structure of the volatile layer will be different. To account for this effect, {\scriptsize BICEPS} uses the non-grey atmosphere model of \citet{Parmentier15} to calculate the temperature profile of the outermost volatile layer. We further use the Schwarzschild criterion to determine if at a certain depth the volatile layer is stable against convection, i.e. if the temperature gradient is radiative or adiabatic \citep{Kippen12}.
At last the transit radius is calculated by determining the radius at the chord optical depth of $\tau_\text{chord}=2/3$ \citep{Guillot10,Parmentier14}. The opacities used in this calculation are taken from \citet{Freedman14}. For more details on the internal structure model, especially the used equations of state, we refer to \citet{haldemann_2023b}. 


\subsection{Model limitations}\label{app_model_limitations}
Several simplifications affect the results of our formation-evolution-structure calculations.
\begin{enumerate}
    \item Solar elemental abundances:
    \begin{enumerate}
         \item The rocky component of the planets is always assumed to be Earth-like in composition, i.e. with 33\% iron and 67\% silicates by mass. In reality, a spread in iron/silicate abundances is expected for stars that have  elemental abundances different than solar. The minimum iron mass fraction expected for stars in the thin-disc (where the Sun and most planet-hosting stars are) is of $\sim20\%$, \citep[][their fig.A.1 ]{Michel20}. 
        
        \item The ice-to-rock ratio of 1:1 just beyond the iceline also stems from solar abundances. Stars with other elemental abundances are expected to exhibit different ratios. For thin-disc stars, the minimum water mass fraction of solid material beyond the iceline is of $\sim30\%$ \citep[][their fig.A.2]{Michel20}.  
    \end{enumerate}
    These two compositional assumptions imply that the "depleted diagonal band" found in the M-R diagram of Fig.\ref{Fig_MRfice} should still exist, but could be narrower, between lines of 20\% iron mass fraction for rocky composition, and of 30\% water mass fraction. This would also translate in a persistent density valley for all stellar types between the same compositional lines.

    \item Only one planetary seed per disc: N-body interactions are not yet included in {\scriptsize PLANETALP}. This implies that the planet typically reaches the disc inner edge when migrating inwards, and the possibility of being trapped in resonant chains with other planets at different semi-major axes cannot be modelled. This model limitation has a stark impact on the planets' final orbital period, which should not be directly compared to observations. 

    \item Planetary seeds are all placed at the beginning of the simulations (time = 0). This also affects the final orbital period of wet vs. rocky planets. Formation time-scales are shorter for wet planets (as we showed in V20), and as a consequence, they reach the inner disc at early times, when the zero torque location -the planet migration trap- is located near the inner edge of the disc. As time evolves, this migration trap location moves slightly outwards and dry planets tend to be trapped at larger periods. This is why in Fig.\ref{fig:Rp_Porb}, most of the wet planets tend to be located at orbital periods of $\sim12$ days, while the dry ones are mostly located at orbital periods $\gtrsim 13$ days. 

    \item Planetary cores are assumed to grow only due to pebbles. In reality, pebbles are converting into planetesimals in the disc \citep{Drazkowska17, Lenz19}, and planetesimals could also contribute to the core growth. Except very few cases \citep{A18,Guilera20,Kessler23}, formation models tend to assume only one of the two solid accretion types due to the uncertain pebble-to-planetesimal ratios along the disc. Planetesimal accretion would yield more mixed compositions \citep{Brugger20, Burn21}, blurring to some extent the density valley. 

  \item The atmospheric escape rates for atmospheres enriched in water are taken from empirical photoevaporation laws from protoplanetary discs \citep{Ercolano10} (Sect.\ref{App_evol}). Detailed hydrodynamical atmospheric escape simulations for mixtures of H, He and H$_2$O should be developed in the future.
    
    \item Simplifications in the internal structure calculations: our upper irradiated atmospheres assume a solar composition \citep{Parmentier15}. In addition, the gas opacities in the radiative zones are calculated with the model of \citet{Freedman14}, which accounts for certain degree of envelope enrichment, but are not specific to water.
    
\end{enumerate}

\section{Extended results}
\subsection{Bare cores after disc dissipation}
The effect found in V20 is in principle expected to occur for any stellar mass, because it is related to the change of the pebbles' properties at the water iceline. 
This, in turns, produces more massive icy cores than rocky ones, and this effects as well the core sizes.

\begin{figure*}
   \centering
   \includegraphics[width=\linewidth]{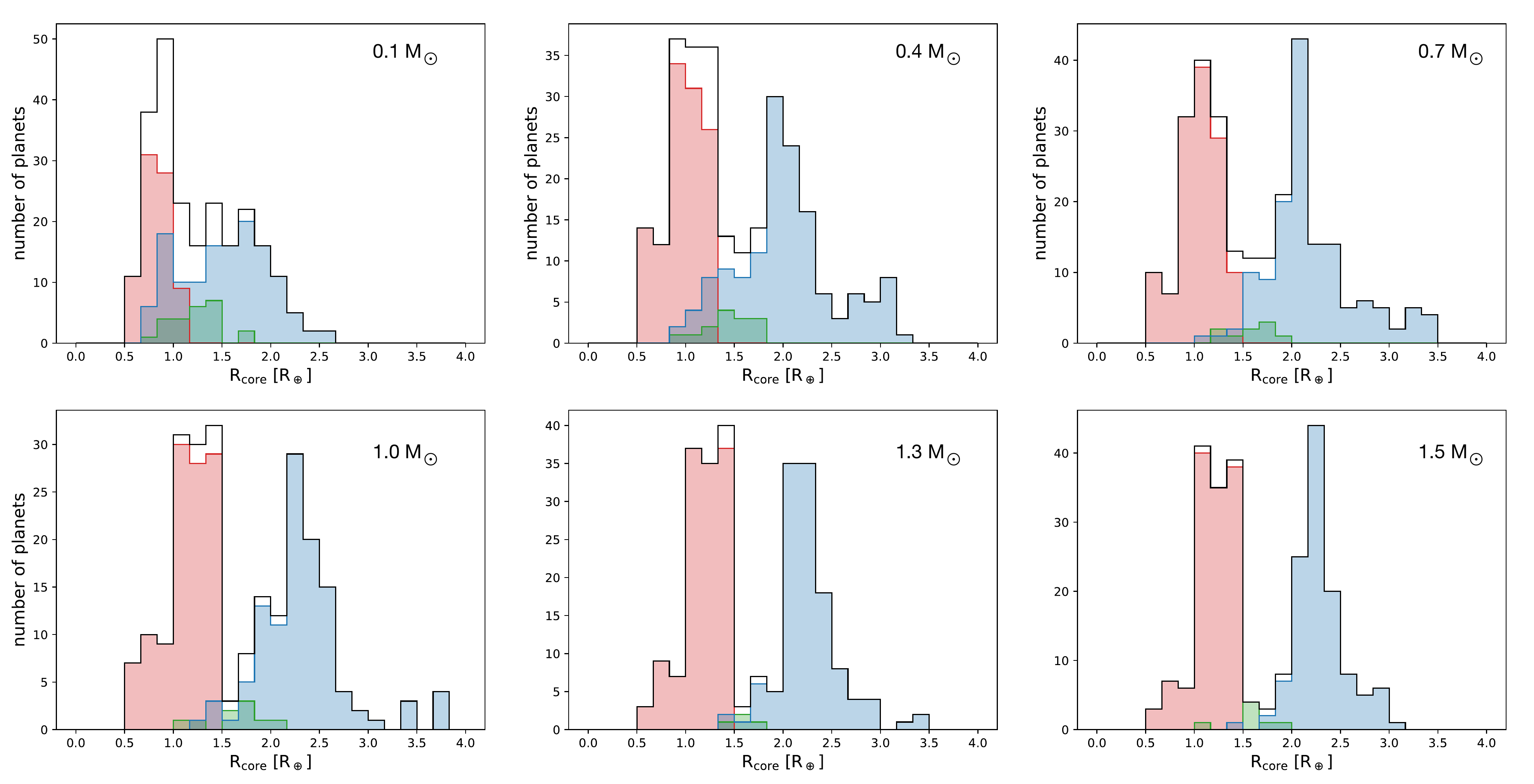}
   \caption{Histograms of core sizes at the end of formation, for the different stellar masses, and for planets with final orbital period below 100 days. Red bars indicate $f_{\rm H_2O,core}<5\%$, green bars $5\% \leq f_{\rm H_2O,core}<45\%$, and blue bars $f_{\rm H_2O,core} \geq 45\%$, where $f_{\rm H_2O,core}$ is the core water mass fraction at the end of formation. The black lines show the overall core size distribution. }
              \label{Fig_histRcores}%
    \end{figure*}

In fig. \ref{Fig_histRcores} we show the histograms of the core sizes of the simulated planets at the end of formation. The blue bars indicate a  core water mass fraction larger than $f_{\rm H_2O,core} > 45\%$, the red ones, dry cores ($f_{\rm H_2O,core} < 5\%$), and the green bars intermediate, mixed compositions ($5\% \leq f_{\rm H_2O,core} \leq 45\%$).
We note that, as expected, icy cores tend to be bigger than rocky cores, for all stellar masses. A deficit of cores with sizes in the range of $\sim 1.5-2$ \RE  is clear for $M_{\star} \geq 0.4 \, \Msun$. The valley separates dry from wet cores for $M_{\star} \geq 0.4 \, \Msun$.
Interestingly, the lower the stellar mass, the larger the overlap in size between the dry and wet populations, and the larger the number of planets with intermediate compositions. In particular, for $M_{\star} = 0.1 \, \Msun$ there is no second peak, as many icy cores are very small. This behaviour is due to migration. Type-I migration occurs for lower planet mass at lower stellar masses \citep{Paardekooper2011}. This allows for smaller icy planets to reach the inner system for the low stellar mass cases. This feature was also reported by \citet{Burn21}.
\begin{figure*}[t!]
   \centering
   \includegraphics[angle=0,width=1.0\linewidth]{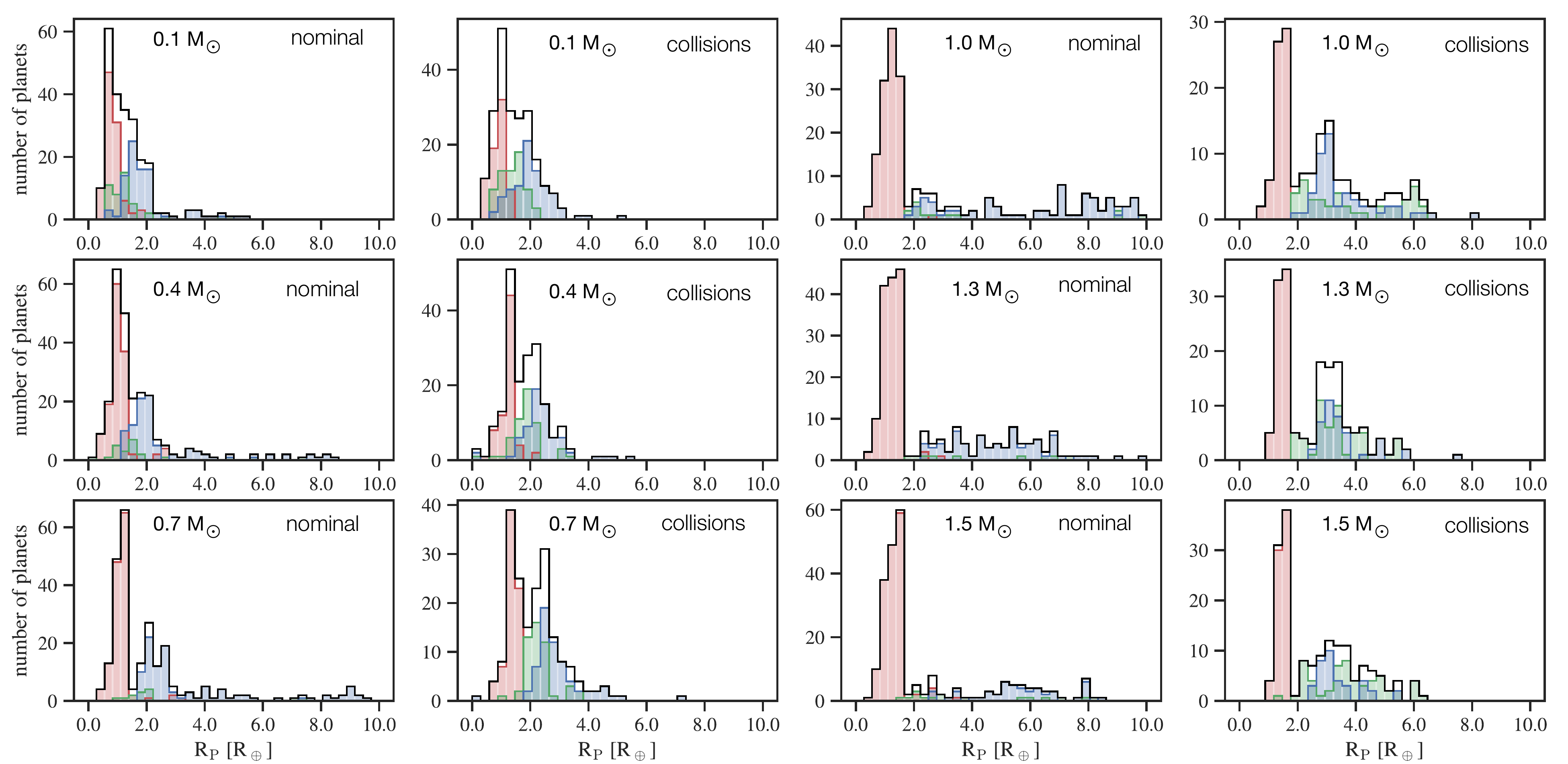} 
   \caption{Histograms of planets' radii after 2 Gyr of evolution, for different stellar masses, and for planets with final orbital period below 100 days. Red bars indicate $f^{'}_{\rm H_2O,core}<5\%$, green bars $5\% \leq f^{'}_{\rm H_2O,core}<45\%$, and blue bars $f^{'}_{\rm H_2O,core} \geq 45\%$, where $f^{'}_{\rm H_2O,core}$ is the core water mass fraction at 2 Gyr of evolution. The black lines show the overall planet size distribution. Left panels display results for the nominal set-up, and right panels for the collisional set-up (see Sect.\ref{App_collisions}).}
              \label{Fig_Ap2}%
    \end{figure*}
\subsection{After 2 Gyr of evolution}
Figure \ref{Fig_Ap2} shows the histograms of the planet radii at 2 Gyr of evolution, for both the nominal and collisional cases.
In agreement with Fig.\ref{Fig_histRcores}, the radius valley is non-existent for $M_{\star} = 0.1 \, \Msun$, and starts to be visible for $M_{\star} \geq 0.4 \, \Msun$ for the collisional case (which resembles better the bare core cases of Fig.\ref{Fig_histRcores} since more atmospheric mass is removed when we allow for giant impacts). Interestingly, our model predicts that the case of $M_{\star} = 0.7 \, \Msun$ is the one for which the radius valley is the most prominent, with a clear peak of water-rich mini-Neptunes for both nominal and collisional cases. This is in line with the findings of \citet{Kunimoto20} which report the occurrence rate of mini-Neptunes increasing for k-dwarfs compared to G-dwarfs.
It is important to clarify that these histograms should not be taken as absolute occurrence rates to compare directly with observations. Indeed, this study is not a population synthesis, in the sense that the initial conditions are not taken with the weights of the observed distributions of disc lifetimes. We do consider the ranges of possible initial conditions stemming from observations, but not the weights of the distributions. Nevertheless, the fractions of certain types of planets (e.g, mini-Neptunes) can be compared \textit{relatively}, between our different stellar masses. Our planet formation and evolution parameter study predicts that mini-Neptunes reach their peak of occurrence for stellar masses of $ \Mstar \sim 0.7 \, \Msun$ (strictly speaking, $0.4 \leq  M_{\star} \leq 1.0 \, \Msun$ due to our binning in stellar mass). For the nominal cases of $M_{\star} \gtrsim 1.0 \, \Msun$, the water-rich planets do not concentrate anymore around a peak value, but rather spread over a large range of sizes. This is mainly the effect of gas accretion onto cores that are more massive compared to $M_{\star} \gtrsim 1.0 \, \Msun$ and which can bind larger amounts of gas. When gas is removed by collisions, the mini-Neptune peak re-emerges, similarly as we found in V20. Another effect removing mini-Neptunes for the case of the most massive stars ($M_{\star} \gtrsim 1.3 \, \Msun$) is the strong photoevaporation, which removes completely the H/He/H$_2$O envelopes for more planets compared to $M_{\star}=1.0 \, \Msun$. Hence, planets that were born beyond the iceline as ice-rich, can become bare rocky cores by evolution. In our simulations this happens for 28\% of the original water-rich mini-Neptunes for $M_{\star} = 1.5 \, \Msun$ and for 14\% of the original water-rich mini-Neptunes for $M_{\star} = 1.3 \, \Msun$.

\subsection{Water-worlds around low-mass stars migrated from beyond the iceline}\label{app_iceline}
\begin{figure}[ht]
   \centering
   \includegraphics[angle=0,width=1.0\columnwidth]{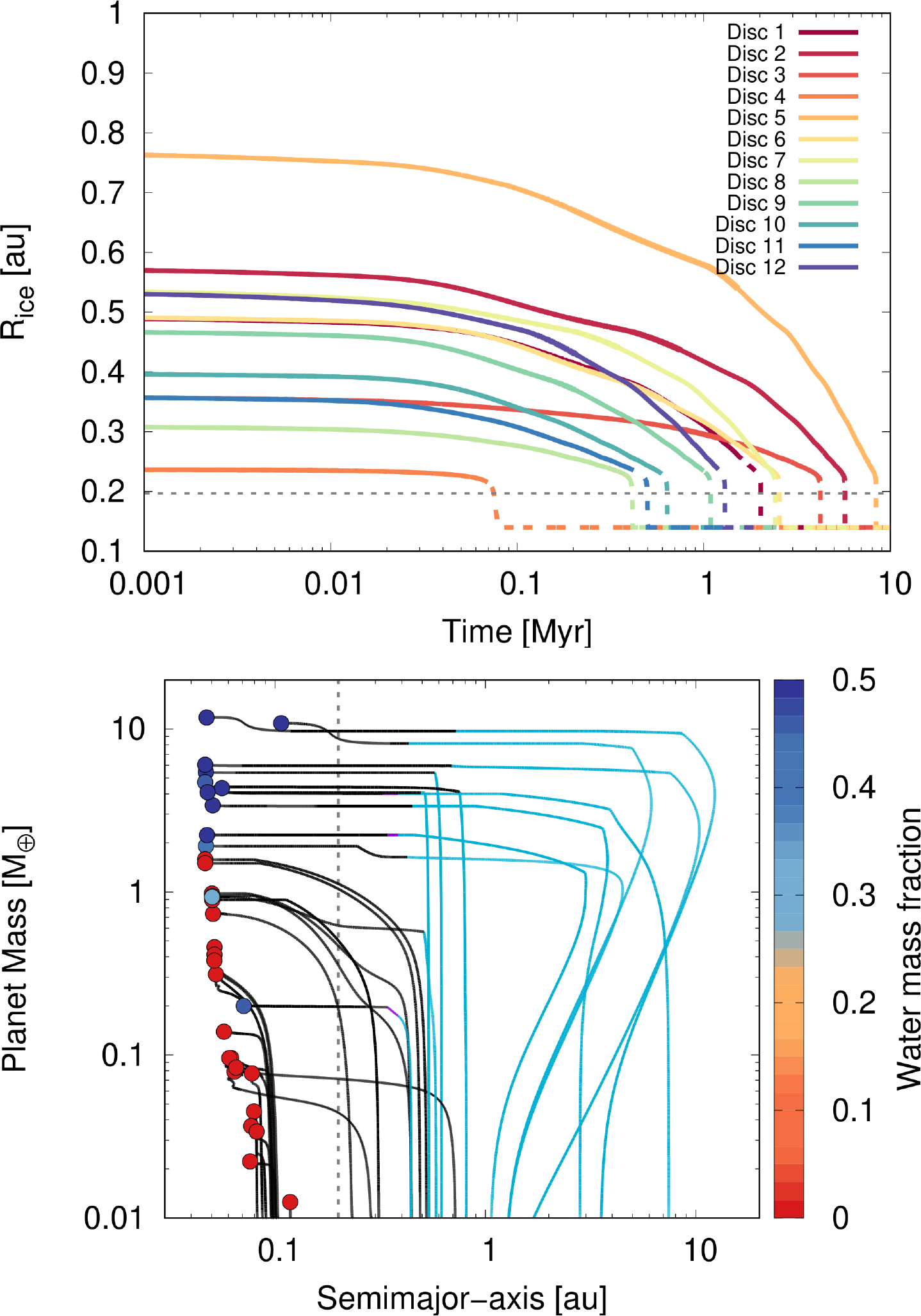}
   \caption{\textit{Top-panel}: time evolution of the icelines for the discs around $0.1M_\odot$ for the different protoplanetary discs considered in our simulations (see Sect.{\ref{App_IniCond}}, for $\alpha=10^{-4}$). The icelines turn from solid to dashed lines when the mid-gap opens in the protoplanetary disc (i.e, when the disc dissipates at the iceline's location). The horizontal grey dashed line represents the orbit of $\sim100$~days. \textit{Bottom-panel:} growth tracks of planets forming around $0.1M_\odot$ that end within the 100-day period. Black lines indicate that the planet is located inside the iceline, while torquoise indicate the planet is beyond the iceline. Clearly, all planets which accreted some ice started to form beyond a 100-day period.}
   \label{figApIcelines}
\end{figure}

The iceline of protoplanetary discs around low-mass stars is much closer to the central star than for more massive stars. For example, while for $\Mstar=1.0 \, \Msun$, the iceline typically locates initially at $\sim$2-3 au, for $\Mstar=0.1 \Msun$ it is at$\sim$0.5 au at the beginning of the simulations. Because of this, the existence of water-worlds around low-mass stars could in principle be explained without type-I migration, in a scenario where wet-planets with orbital periods below 100 days simply formed in-situ with the iceline closer in. However, this never happens in our simulations, water-worlds always accrete the bulk of their ices beyond the iceline and migrate inwards afterwards. To illustrate this, we show in the left panel of fig. \ref{figApIcelines} that the icelines in discs around $\Mstar=0.1 \Msun$ never manage to cross the 100-day orbit threshold (marked with the horizontal grey dashed line at $\sim0.2$~au) before photoevaporation opens a gap in the disc (time from which the solid lines become dashed lines). 
The right panel of fig. \ref{figApIcelines} shows that planets forming within the iceline for $\Mstar=0.1 \Msun$ are always dry (planetary tracks represented with black curves only), while those that start their growth outside the iceline but end inside it are water-rich (planetary tracks represented by turquoise curves that then turn into black once they cross the iceline) and can only reach the regions of orbital period below 100 days (marked by the dashed vertical line) through type I migration. We have checked that this scenario always takes place with all the icy planets around all the stars considered. For simplicity we only show it for some of them, selected randomly, around $0.1M_\odot$.
\begin{figure*}[ht]
   \centering
   \includegraphics[width=0.99\linewidth]{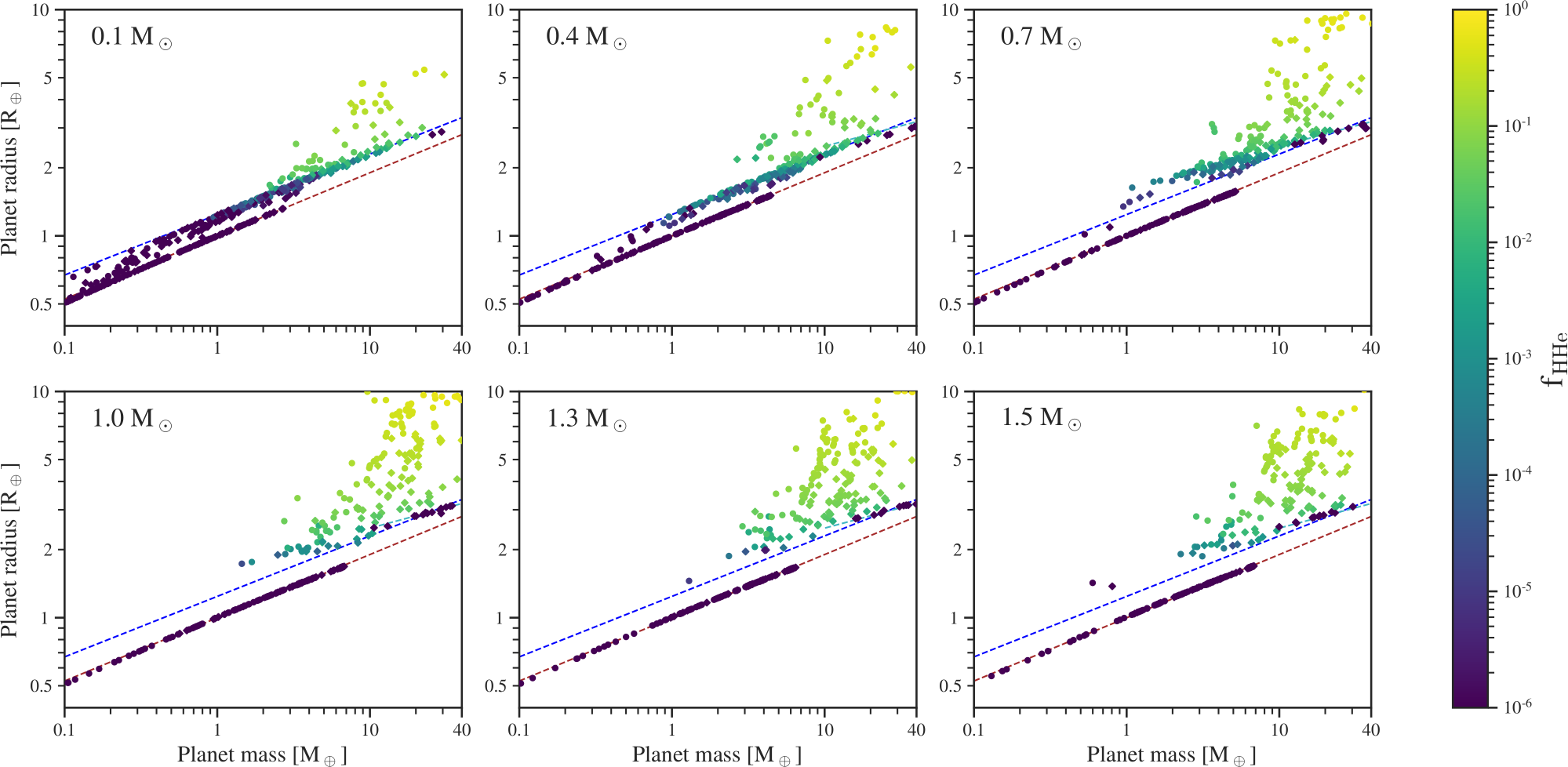}
   \caption{Same as Fig.\ref{Fig_MRfice} but the colour-bar represents the planets' mass fraction of H/He.}
              \label{Fig_MRfhhe}%
\end{figure*}

\subsection{Steam atmospheres}\label{app_steam}
\begin{figure}[ht]
   \sidecaption
   \includegraphics[width=1\linewidth]{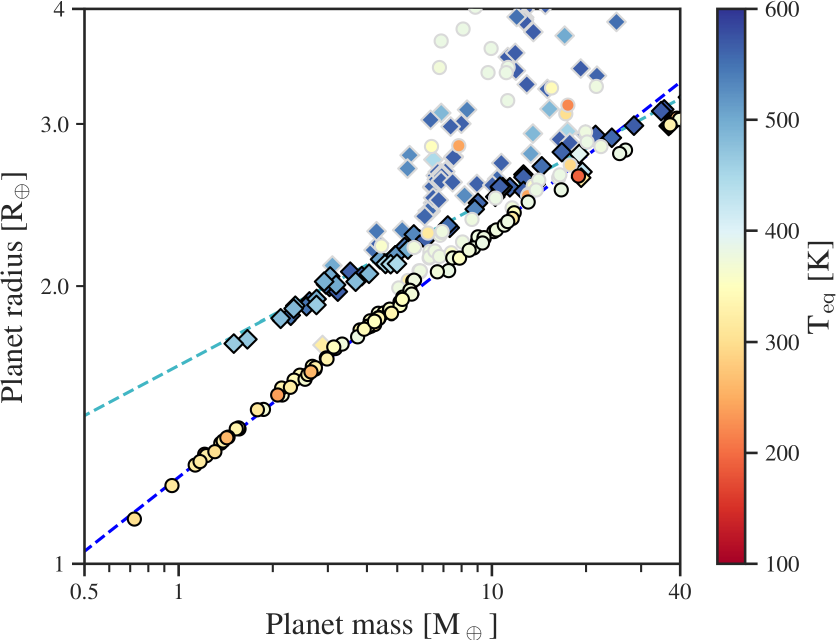}
   \caption{Mass-Radius diagram of water-rich planets ($\fw\ge 0.45$) around $\Mstar=~0.4 \, \Msun$ (circles) and $\Mstar=~0.7 \, \Msun$ (diamonds). Black-bordered symbols indicate $\fhhe\le 10^{-2}$, while grey, $\fhhe>10^{-2}$. The colour-bar is the equilibrium temperature of the planets at zero albedo. The dashed-blue line shows the condensed water line, the light-blue dashed a fitting M-R relation to the steam worlds, described in Sect.\ref{app_MRfits}.}
              \label{Fig_ApB4}%
    \end{figure}

An interesting feature of Fig.\ref{Fig_MRfice} is that for $\Mstar\geq 0.7 \, \Msun$, most of the water-rich, H/He-poor planets  ($\fw>0.45$ and $\fhhe<10^{-2}$, see Figs.\ref{Fig_MRfice} and \ref{Fig_MRfhhe}) lie above the condensed water-line. 
To understand what is puffing the water-rich planets up, first we need to isolate the effect of H/He from the effect of temperature on the planetary radii. For this, we plot in Fig.\ref{Fig_ApB4} the M-R diagram as a function of equilibrium temperature for $\Mstar=0.4.- 0.7 \,\Msun$. The planets are also distinguished by a H/He mass fraction smaller or larger than 0.01 (we refer to them as H/He poor if $\fhhe<10^{-2}$ and H/He-rich otherwise). Three things are clear from this diagram:
\begin{enumerate}
\item Water-rich and H/He-poor planets with the lowest equilibrium temperature (below 400 K, the ones around $\Mstar=0.4\,\Msun$) lie on the condensed water line. These planets are cold enough for water to be in condensed form.
\item All of the water-rich, H/He-poor planets with $\Teq\geq400$ K lie on a different line (light blue in the figure), well described as $\Rp (\Mp)$= $a \, M_{\rm P}^{b}$, with $a=1.64183$ and $b= 0.180572$ (see Sect.\ref{app_MRfits}).
\item All the planets with $\fhhe \gtrsim 10^{-2}$ are scattered either above the light-blue M-R curve for $\Mstar=0.7 \,\Msun$, or above the condensed water line for $\Mstar=0.4 \,\Msun$.
\end{enumerate}
Indeed, for all the planets with $\fhhe>10^{-2}$, the radius is strongly affected by the amount of primordial gas and by the equilibrium temperature. On the other hand, the planets that follow the light-blue line are H/He poor but their full atmospheres are hot enough for water to be present as vapour. 
This is better illustrated in Fig.\ref{fig_PTwaterWorlds} where we show specific pressure-temperature atmospheric profiles (App.\ref{AppPT}).

\subsection{M-R relations: analytical fits}\label{app_MRfits}
In this section, we specify the Mass-Radius analytical curves deployed in Fig.\ref{Fig_MRfice}. Such M-R relations were obtained by fitting our results for the rocky, condensed-water-, and steam-planets computed with BICEPS (Sect.\ref{App_biceps}.)

For rocky planets, defined as those planets with final water contents lower than 5\% in mass and H-He envelopes lower than $10^{-6}$ respect to the total mass, we find the following relation: 
\begin{eqnarray}
\dfrac{\text{R}_{\text{P}}}{\text{R}_{\oplus}}= a_{\text{rocky}} * \left(\dfrac{\text{M}_{\text{P}}}{\text{M}_{\oplus}}\right)^{b_{\text{rocky}}},
\label{eq:fit_rocky_cores}
\end{eqnarray}
with $a_{\text{rocky}}= 0.999009 \pm 0.0002802$, and $b_{\text{rocky}}= 0.279514 \pm 0.0002377$. Our fit is in very good agreement with the one found by \citet{Zeng19}, who found $a_{\text{rocky}}= 1$, and $b_{\text{rocky}}= 1/3.7 \approx 0.27027$.  

For condensed-water planets, which are those planets with final water contents greater than 45\% in mass, H-He envelopes lower than $10^{-2}$ by mass, and $T_{\text{eq}} < 400$~K, we find:
\begin{eqnarray}
\dfrac{\text{R}_{\text{P}}}{\text{R}_{\oplus}}= a_{\text{cw}} * \left(\dfrac{\text{M}_{\text{P}}}{\text{M}_{\oplus}}\right)^{b_{\text{cw}}},
\label{eq:fit_icy_cores}
\end{eqnarray}
with $a_{\text{cw}}= 1.24191 \pm 0.001667$, and $b_{\text{cw}}= 0.267404 \pm 0.0008558$. We note again that our fit is in very good agreement with the one found by \citet{Zeng19} for those planets with 50\% of water by mass, who found $a_{\text{cw}}= 1.24$, and $b_{\text{cw}}= 1/3.7 \approx 0.27027$. 

Finally, for the steam-worlds, defined as the condensed-water worlds except that $T_{\text{eq}}~\ge~400$~K, we obtain the following fit:
\begin{eqnarray}
\dfrac{\text{R}_{\text{P}}}{\text{R}_{\oplus}}= a_{\text{steam}} * \left(\dfrac{\text{M}_{\text{P}}}{\text{M}_{\oplus}}\right)^{b_{\text{steam}}},
\label{eq:fit_steams_planets}
\end{eqnarray}
with $a_{\text{steam}}= 1.64183 \pm 0.008192$, and $b_{\text{steam}}= 0.180572 \pm 0.001608$. 
We note that our calculations indicate an increase of planet radius of 30\%, 25\% and 10\% between steam-worlds and condensed-water-worlds of masses of 1, 2 and 5 \ME, respectively (for planets with water mass fractions of 50\%).
As an interesting final remark, we note that the M-R relation that we find for steam worlds does not depend strongly on equilibrium temperature for different stellar masses. 

\subsection{Mean density: histograms}\label{app_density}
\begin{figure*}[t!]
   \centering
   \includegraphics[angle=0,width=1.0\textwidth]{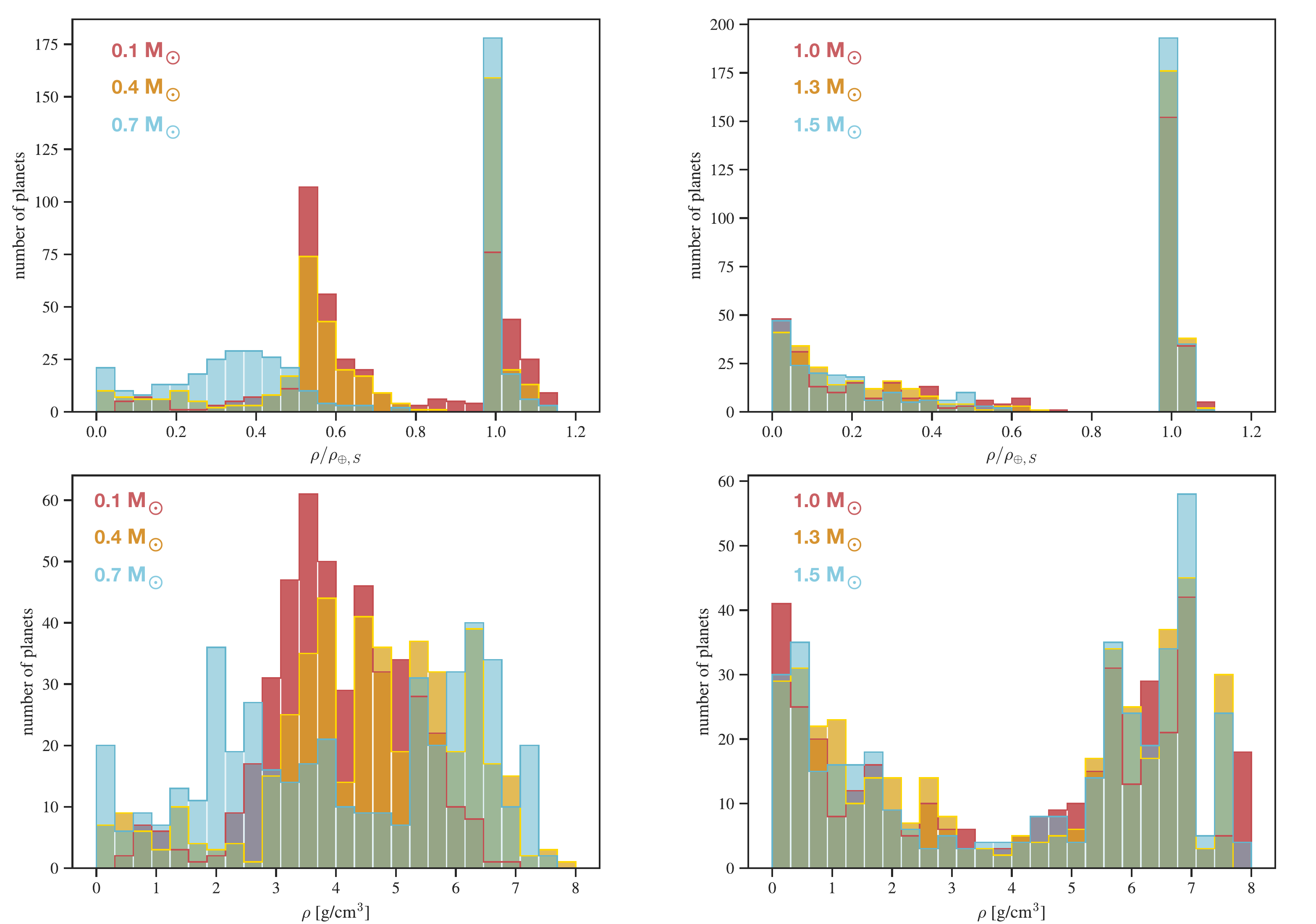}
   \caption{Planetary mean density for the different stellar masses. Left (right) panels show values for the low (high)-mass stars, with the each color corresponding to one stellar mass as indicated in the legend. The top panels show the normalised density (see text), while the bottom panels show the density in cgs.}
   \label{Fig_Ap_histosDensity}
\end{figure*}

Figure \ref{Fig_Ap_histosDensity} shows the histograms of the planetary mean densities for the different stellar masses. The left panels are for the planets around the M- and K-dwarfs, while the right panel for planets orbiting G-F- and A-stars. For the top panels the density is the "normalised density", i.e, the mean density divided by the density that the planet would have if it had an Earth-like composition (i.e, a mass radius-relation given by Eq.\ref{eq:fit_rocky_cores}). The bottom panels display the histograms of the physical density in cgs. Clearly, the density valley is better visualised in terms of normalised density, especially for the low-mass stars. The peak of planets at 0.5-0.6 $\rho/\rho_{\oplus,S}$ of the left-top panel corresponds to the condensed-water worlds, which do not exits in our results for high-mass stars due to the higher equilibrium temperatures (see Fig.\ref{fig_discussion}). In terms of phyiscal density (bottom panels, the valley occurs at $\rho \approx$~4-5 g/cm$^3$ for M- and K-dwarfs, and for $\rho \approx$~3-4 g/cm$^3$ for G-F-A spectral types.) 

\subsection{A few dry mini-Neptunes in the M-R diagram}\label{App_red-outliers}
Figure \ref{Fig_MRfice} contains a few dry "outliers" in the region of the M-R diagram above the condensed water-line, where most planets are water-rich. These planets were actually formed beyond the iceline, just like the wet planets that surround them. The difference with the latter is that they lost most of their water (and H/He) due to photoevaporation, during the evolutionary phase. 
To illustrate this, we show in Table \ref{table_2planets}, the initial and final planet mass, water mass and H/He mass for two planets that end up very close in the M-R diagram of $\Mstar=1.5 \, \Msun$. We also tabulate the rock mass (invariable during the evolution), and the initial and final water mass fraction ($\fw$), H/He mass fraction ($\fhhe$) and envelope metallicity ($\Zenv$, defined as the mass of water divided by the mass of water plus H/He). One of the planets ends up water-rich and the other water-poor after 2 Gyr of evolution (referred as ``wet" and ``dry" planets in the table, respectively).
 The reason behind the different evolutionary paths is the atmospheric escape rate going as $\sim \Zenv^{-0.77}$ (with $\Zenv$ being the envelope water mass fraction, which remains constant throughout the evolution by hypothesis). This means that the smaller the initial $\Zenv$, the faster the atmospheric mass loss. Indeed, the two planets have very different initial $\Zenv$, despite of having similar initial $\fw$. The smaller $\Zenv$ of the dry planet is a consequence of having a much larger initial amount of H/He. Because of the smaller $\Zenv$ of the dry planet, is loses 99.45\% of its initial envelope during 2 Gyr of evolution, while the "wet" planet loses only 6.46\% of it. This leaves the "dry" planet with only $\fw = 0.55\%$ of bulk water fraction at the end of the evolutionary phase, while the "wet" planet keeps $\fw = 47.6\%$. Despite of being highly "desiccated", the dry planet would still show a high water signature in its atmospheric spectra, due to keeping an atmospheric water mass fraction (or envelope metallicity) of $\Zenv \approx 55\%$. We caution, however, that atmospheric escape rates of envelopes enriched with elements heavier that H/He are highly uncertain, and that much theoretical research is needed to provide meaningful predictions of water mass fraction during the evolutionary phase.
\begin{table}[h]
   \begin{center}
\begin{tabular}{ p{1.9cm} p{2.45cm}  | p{2.45cm} }
  \hline
  \hline
  &   dry planet & wet planet \\
 \end{tabular}
 
 \begin{tabular}{ |p{1.9cm}||p{1cm} p{1cm} | p{1cm} p{1cm}|}
\hline
 Mass & Initial & Final & Initial & Final\\
 \hline
       $\Mp$  [\ME] &  13.604  & 4.855 &  4.833 & 4.674  \\
       $\MHHe$  [\ME] & 4.008   &0.02218  & 0.1519  & 0.142  \\
       $\Mw$  [\ME]  & 4.791  & 0.02651  & 2.304   & 2.155\\
       M$_{\rm rock}$ & 4.806 & 4.806	& 2.377 & 2.377  \\
       $\fw$ & 0.499 & 0.00550 & 0.4922 & 0.4755\\
       $\fhhe$ & 0.2946 & 0.00457 & 0.03282 & 0.0304\\
       $\Zenv$ & 0.5445 & 0.5445  & 0.9382 & 0.9382 \\       
\hline
    \end{tabular}
    \caption{Physical quantities of two planets around $\Mstar = 1.5 \,\Msun$ that finish with very similar mass and radius but with very different compositions. "Initial" and "final" refer to the onset and end of the evolutionary phase, respectively.The mass of rocks remains invariant during the evolution.}
    \label{table_2planets}
    \end{center}
\end{table}    

\section{Extended discussions}
\subsection{Dependence of the radius valley location on the formation model parameters}

Planet formation models contain a set of free parameters, with a range of possible values inferred from either observations or lab experiments. 
These parameters can affect the outcome of planet formation simulations in regard to the predicted planet mass, planet radius, and orbital distance to the star, among others. 
The prime parameters that can result in very different outcomes in our pebble-based formation model are: 
\begin{enumerate}
    \item the $\alpha$-viscosity parameter
    \item the fragmentation threshold velocity of pebbles
    \item the opacity of the gaseous envelope
\end{enumerate}
While a deep analysis of the impact of these parameters on our results is beyond the scope of the present work, it is nevertheless pertinent to discuss the effect they could have on the location of the radius valley. 

\subsubsection{Dependence on the $\alpha$-viscosity parameter}
The $\alpha$-viscosity parameter affects the growth of the planetary cores in two ways. On the one hand, it affects the pebbles' sizes (the higher the $\alpha$, the more turbulent is the disc, hence the more the pebbles collide and break, maintaining the pebbles at smaller sizes). The larger the pebbles, the faster the core grows \citep[e.g][]{Lambrechts14, Venturini20SE}. On the other hand, the disc viscosity impacts the disc's structure, including the disc aspect ratio. This means that the pebble isolation mass --the maximum mass a core can grow by pebble accretion--, is indirectly affected by $\alpha$, because the pebble isolation mass depends strongly on the disc aspect ratio.
In principle, the larger the $\alpha$, the larger the pebble isolation mass, because the larger the viscosity, the more massive the planet needs to be to open a partial gap. We checked that the pebble isolation mass at the iceline 
is of about 8 M$_{\oplus}$ for $\alpha=10^{-5}$, compared to $\sim$15 M$_{\oplus}$ for $\alpha=10^{-4}$ for a standard disc around a Sun-like star. A planet of 8 M$_{\oplus}$ with $\fw=0.5$ and without H-He would have a size of 2.86 R$_{\oplus}$ (from using the M-R relation of Eq.\ref{eq:fit_steams_planets}), which means that the mean of water-worlds forming at such low viscosity are still expected to contribute mainly to the peak of mini-Neptunes and not to fill the radius valley (at $\approx 1.8-1.9$ R$_{\oplus}$ for $\Mstar=1.0$ M$_{\odot}$).

In \citet{Venturini20SE}, we discussed extensively the impact of $\alpha$ on the planetary growth by pebble accretion within the iceline, considering values of $\alpha=10^{-5}, 10^{-4} \text{and} 10^{-3}$ \citep[which are the values inferred from disc observations, see e.g][]{Rosotti2023}.
We found that for $\alpha = 10^{-3}$, the rocky pebbles were so small ($\sim10 \, \mu$m) that planets were basically not growing inside the iceline. On the other extreme, for $\alpha = 10^{-5}$, we found that the pebbles reach much bigger sizes ($\sim1$~cm) in the dry inner regions, and planetary growth proceeds extremely fast (in the order of $10^4$ years). 
We also found that for the formation of rocky cores, considering $\alpha = 10^{-5}$ increases the maximum bare rocky core mass from 5 $\text{M}_{\oplus}$ to 7 $\text{M}_\oplus$ -respect to $\alpha = 10^{-4}$. This would still yield a maximum radius of bare rocky cores of 1.69-1.95 $\text{R}_{\oplus}$ (from using the M-R relation of \citep{Zeng19} and of Eq.\ref{eq:fit_rocky_cores}), that is, at the location of the valley.

\begin{figure*}
    \includegraphics[width=1.\textwidth]{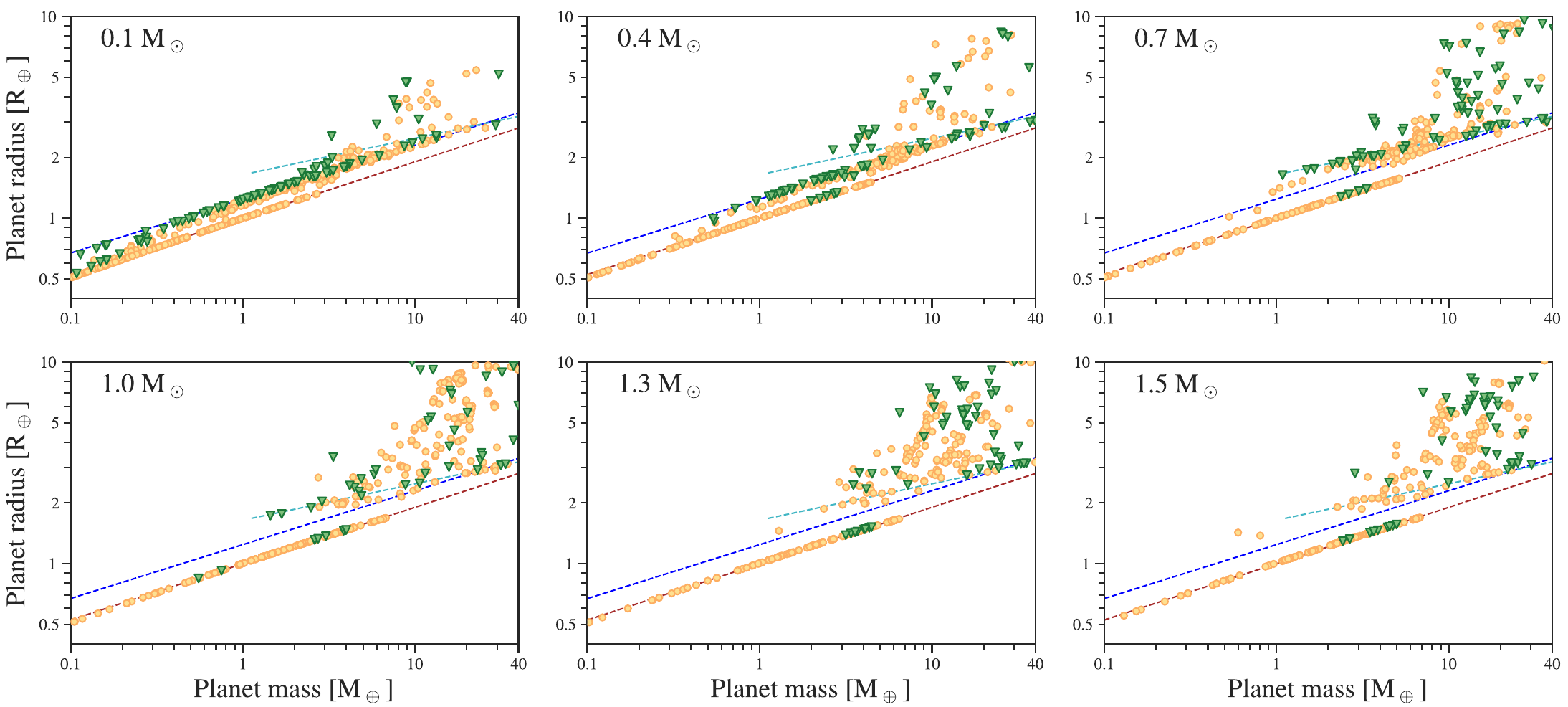}
    \caption{M-R diagrams according to the adopted values of the $\alpha$-viscosity parameter. Green triangles: $\alpha=10^{-3}$, orange circles:$\alpha=10^{-4}$. Both nominal and collisional cases are displayed.}
    \label{MR_alpha}
\end{figure*}

It is also important to mention that the planet migration prescriptions adopted in our model are not really valid for $\alpha\sim10^{-5}$. For such small values of $\alpha$, even a low-mass planet could open a partial gap. \citet{McNally2019} showed that in this case, vortices form at the edges of the partial gap, changing the migration regime. Thus, based on our previous results, in this work we decided to keep the same approach as in V20, using only $\alpha = 10^{-4}$ and $\alpha = 10^{-3}$.
We analyse the effect of this two values of the $\alpha$-viscosity parameter on the location of the radius valley on Fig.\ref{MR_alpha}. The plot shows the M-R diagrams for the synthetic planets, distinguishing according to the cases where $\alpha = 10^{-4}$ (orange circles) or $\alpha = 10^{-3}$ (green triangles). Overall, we notice that the void of the "diagonal band" delimited by the Earth-like composition and the condensed water lines (which sets the location of the radius and density valley, see main text), is preserved for both values of $\alpha$. 
For $\alpha = 10^{-3}$ we have very few rocky planets, which is expected due to the extremely low pebble sizes, resulting in negligible core growth inside the iceline \citep{Venturini20SE}.
Overall, the jump in pebble size at the iceline is the underlying reason for the dichotomy in core size and core composition in our pebble-based formation model \citep{Venturini20Letter}, and this jump exists for any value of the $\alpha$-viscosity parameter \citep[Fig.4 of, top panel of][]{Venturini20SE}.

\subsubsection{Dependence on the fragmentation threshold velocity of pebbles} \label{App_frag_pebbles}
Regarding the impact of the fragmentation threshold velocities on the results, we showed in V20 for Sun-like stars that considering v$_{\rm th}$= 1 m/s for silicates and v$_{\rm th}$= 5 m/s for icy pebbles gives very similar results than our nominal assumption of v$_{\rm th}$= 1 m/s for silicates and v$_{\rm th}$= 10 m/s for icy pebbles. If instead v$_{\rm th}$ = 2 m/s is chosen for the icy pebbles, less planets with the maximum core ice mass fraction of 0.5 form, and more planets with intermediate ice compositions emerge. This would contribute to the filling of the radius valley. Finally, we tested as well a very low fragmentation threshold velocity of v$_{\rm th}$= 1 m/s, constant along the disc. In that case, the pebbles remain so small along the disc, that planet formation by pebble accretion is very inefficient: no planet forms with mass above Earth, and no planet migrates within 100 days of orbital period. 

The values of the fragmentation threshold velocity suited for protoplanetary disc conditions have been a topic of intense debate in the past years. The values adopted in this work (10 m/s beyond the water-iceline and 1m/s within the iceline), are the standard values used in dust growth and evolution studies \citep{Drazkowska16, Drazkowska17}, and on planet formation models, \citep[e.g.][]{Drazkowska2021, Schneider2021, Savvidou2023}. These values are also supported by many experimental works  \citep[e.g.][]{Gundlach15, Musiolik2016}. While recent laboratory experiments found that amorphous water-ice particles could be as fragile as silicates \citep[e.g.][]{Gundlach2018, Musiolik19}, more recent works, such as those of \citet{Nietiadi2020} and \citet{Musiolik2021}, revealed that ices of various volatile molecules on the surface of dust aggregates act like liquids during collisions. This behaviour increases the sticking properties and enables higher-velocity collisions that do not lead to the fragmentation of ice-coated dust aggregates. Thus, given the current knowledge on the topic, we consider that a value of 10 m/s for the fragmentation threshold velocity of ice-rich grains is a sufficiently reasonable assumption. In addition, fragmentation threshold velocities of 1 m/s for silicate and 10 m/s for icy pebbles were recently adopted by \citet{Canas2024} to explain the density dichotomy in the Kuiper Belt Objects. For our results, the location of the radius valley is invariant as long as the fragmentation threshold velocity of icy pebbles is in the range of 5-10 m/s.

\subsubsection{Dependence on the envelope's opacity}\label{App_opac}
The opacities in the gaseous envelope control its cooling and contraction, and hence the gas accretion rate as long as the planet's envelope connects smoothly with the gaseous disc (attached phase). 
The largest uncertainty in the opacity values stem from the poorly constrain grain sizes within a planetary envelope. In principle, grains can settle and grow compared to ISM sizes, which goes in the direction of reducing the dust opacities compared to the ISM. However, the level of turbulence in planetary envelopes is hard to predict, the growth of grains depends upon their size and structure (for instance, fractal versus non-fractal, which is unknown), and as well on the magnitude of the envelopes "recycling". Indeed, recent 3D hydrodynamical simulations show that as long as a planet is located at short orbital distances \citep[within approximately 1 au from the central star][]{Ormel15, Moldenhauer23, Wang-Ormel23}, the gas that is accreted by the planet can flow back to the protoplanetary disc. This means that for planets building the envelopes at short orbital distances (as in our scenario, where planets migrate fast towards the disc inner edge and continue accreting gas at the stranded locations), the small dust grains would be constantly replenishing the envelopes as fresh gas flows in. This is why we chose the high grain opacity values from \citet{BL94}, which correspond to the ISM size grains (and whose values vary with pressure and temperature).

In any case, due to the uncertainty on the dust opacity values of the planetary envelopes during planet formation, it is worth to discuss its impact on the location of the radius valley. The study of \citet{Mordasini20} studies the emergence of the radius valley taking initial conditions from population synthesis studies which assume, contrary to us, a very low dust opacity during the formation phase \citep[][reduced by a factor of 0.003]{BL94}. The location of the valley, corresponding to the maximum mass of bare rocky cores, is at 1.67 R$_\oplus$ at 0.1 au for a Sun-like star \citet[][his fig.4]{Mordasini20}. We obtain the exact same value in our simulations with high dust opacity (nominal case of $\Mstar=1.0 \,\Msun$ displayed in Fig.\ref{Fig_Ap2}). Indeed, \citet{Mordasini20} discusses the practically negligible effect of the opacities on the radius valley location in his Sect.2.3.7. The reason for this is the following. At the end of disc lifetime, the lower the envelope dust opacity, the more massive the gaseous envelope for a given core mass. A more massive envelope is more extended than a thin one, which increases the atmospheric escape rate \citep[which depends on the planet radius to the cube][]{Owen17, Mordasini20}. Thus, even if the planet started with a thicker envelope, photoevaporation will remove it completely for the ranges of core masses we are discussing here (below approximately 10 M$_\oplus$). We found this exact same behaviour in our study of \citet{Venturini20SE} when we tested a set of simulations with 100$\times$ reduced \citet{BL94} opacities: the location of the radius valley was unchanged.

\subsection{Condensed-water- or steam-worlds? Dependence on equilibrium temperature and model assumptions.}\label{AppPT}
\begin{figure}[ht]
    \centering
    \includegraphics[width=1.\columnwidth]{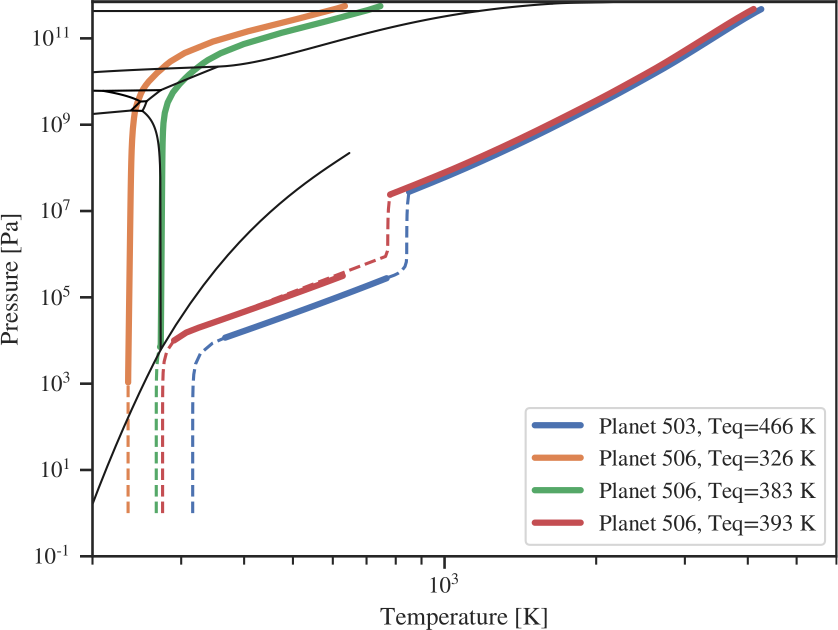}
    \caption{Pressure-temperature profiles of two synthetic mini-Neptunes with $\fw = 0.5$ and equilibrium temperatures of 466 K (planet 503, blue line) and 326 K (planet 506, orange line). The black solid lines in the background are the phase boundaries of pure H$_2$O given by the AQUA EOS \citep{haldemann_2020}. Both planets have a similar mass of approx. 1.5 \ME, but different sizes due to the different physical state of water (planet 503 is a steam-world, with $\Rp =1.690$~\RE and planet 506 is a condensed-water world with $\Rp = 1.657$~\RE). The red and green curves correspond to planet 503, but reducing the equilibrium temperature to see where the transition from steam-to condensed-water world occurs.}
    \label{fig_PTwaterWorlds}
\end{figure}

It is well known that the atmospheric composition, distribution of heavy elements and choice of irradiation model affect the calculation of the planetary radii \citep{Guillot10}. In V20 we argued that assuming the water as mixed with H/He was the most physically motivated assumption since for Sun-like stars all our synthetic planets had temperatures high enough for water to be in the form of steam throughout all the atmospheres. 
In this work, we made the same assumption of water mixed with H/He during the evolution simulations and the computation of the planetary radii. However, we noted that our synthetic planets can now have temperatures low enough for water to condense, particularly around M-dwarfs (our $\Mstar=0.1, 0.4 \, \Msun$ cases, App.\ref{app_steam}). The fate of condensed water in H$_{2}$O-H-He atmosphere is not trivial to predict. In principle, condensed water on the top layers would settle into deeper regions and if there is atmospheric circulation, would rise up again. Thus, the assumption of uniform envelope metallicity might still hold. For the case of pure water envelopes, a better physical treatment would be to allow water to sink to the next layer, as long as the condensation conditions in pressure and temperature hold. Unfortunately, our structure code cannot handle yet compositional gradients properly. However, we caution that changing the distribution of water will probably change the value of planet radius in cases where water condensation takes place.  
Not only the distribution of water affects the resulting transiting radius, but also other model assumptions such as the choice of opacities and the irradiation model. 
We note that, compared to our results, the models of \citet{Aguichine21} predict a substantially larger planet radius for planets composed of half-rocks-half-water by mass when $\Teq \geq 400 K$ (their Fig. 5). The key difference with our model is their assumption of the core-envelope boundary to be extremely hot (in a region of the phase diagram of water where water is in 'supercritical' state). This assumption is based on evolution calculations accounting for the 'runaway greenhouse effect of water' \citep[e.g]{Turbet19}. More recent work points out that such effect might have been overestimated in the past due to the assumption of purely convective envelopes \citep{Selsis23}.

\begin{figure*}[ht]
   \centering
   \includegraphics[angle=0,width=0.91\textwidth]{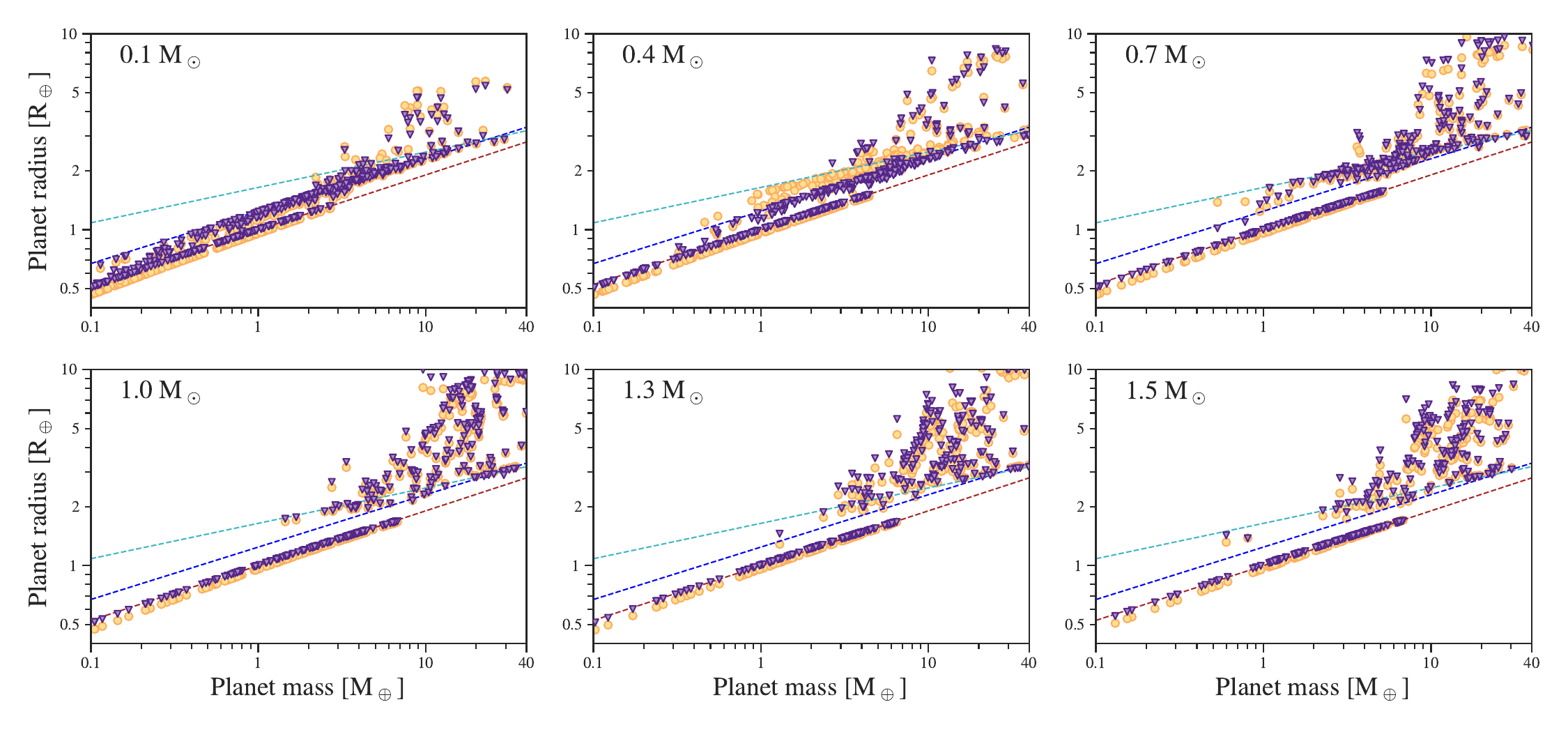}
   \caption{M-R diagrams comparing planetary radii obtained with {\scriptsize BICEPS} (purple triangles) vs. {\scriptsize COMPLETO} (orange circles). For both set-ups water is assumed to be mixed uniformly with the H/He.}
   \label{figApB5}
\end{figure*}

In our calculations we do not impose any condition on the pressure-temperature of the envelope-core boundary, we simply integrate the structure equations inwards from the top of the atmosphere (with the outer boundary conditions depending on the equilibrium temperature, and on the luminosity that the planet irradiates according to its cooling history, see App.\ref{App_evol}). The type of P-T profiles that we find for planets with pure water envelopes are illustrated in Fig.\ref{fig_PTwaterWorlds}. In the figure, we show the profiles of two planets, one of the two has a fully condensed water envelope while the other a full steam one. To better capture the transition between the two, we run again the case of the steam world but reducing its equilibrium temperature (red and green curves of Fig.\ref{fig_PTwaterWorlds}). We note that the transition between the condensed and steam worlds happens for $383<\Teq <396 K$, which in out model translates into a temperature at the top of the atmosphere of $268<T_{\rm out}<276~$K, respectively. 
The reason for $T_{\rm out}$ being smaller that $\Teq $ by $\sim80-120~$K is the non-grey atmospheric model of \citet{Parmentier15}. The previous grey atmospheric model of \citet{Guillot10} also yields $T_{\rm out}$ smaller than $\Teq$, but only by $\sim20~$K. The nearly 100 K difference in Tout between the 2 atmospheric models makes the water of the planets orbiting $\Mstar= 0.4 \, \Msun$ be in condensed form when using the non-grey irradiated model of \citet{Parmentier15} (used in {\scriptsize COMPLETO}) compared to vapour form when using the one of \citet{Guillot10} (used in {\scriptsize COMPLETO}). This is shown in Fig.\ref{figApB5}.
To summarise, the radii of water-worlds near the condensation line are very sensitive to model assumptions. More efforts are needed on the modelling of the atmospheres of water-worlds to compute accurate transit radii.

\begin{figure*}[h!]
    \centering
    \includegraphics[width=0.92\linewidth]{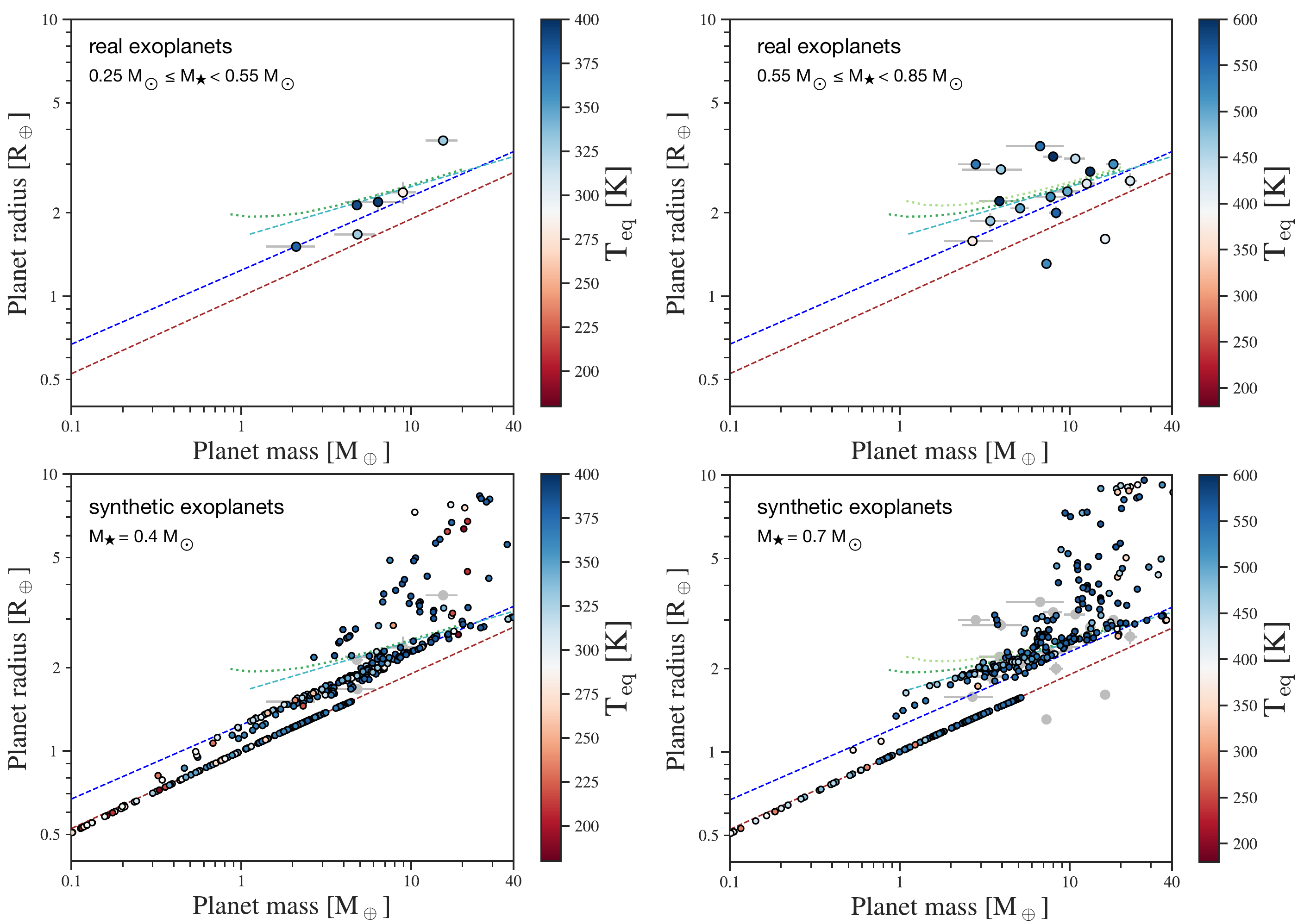}
    \caption{Comparison of real exoplanets for $\Mstar=~0.4 \, \Msun$  (left panels) and $\Mstar=~0.7 \, \Msun$ (right panels). The limit in equilibrium temperature is for proper comparison with our simulations. The brown- and blue-dashed lines correspond to Earth-like and 50\% water-50\% Earth-like compositions, with dark-blue indicating condensed-water and light-blue, steam-worlds (as defined in Ap.\ref{app_MRfits}). The green-dotted curves correspond to 50\% steam-50\% Earth-like from \citet{Aguichine21} ($\Teq=400$ K for dark-green and $\Teq=600$ K for light-green). The real exoplanets are those with a relative error in radius less than 25\% and in mass of less than 70\%, with the data taken form the NASA Exoplanet Archive on 08.09.23. For the bottom panels, the real exoplanets are shown in the background in grey.}
    \label{fig_real_synt}
\end{figure*}

\subsection{Where are the steam-worlds?}\label{App_real_vs_synt}
A controversial aspect of the \citet{Luque22} study is the conclusion that the planets that fall on the condensed water line in the Mass-Radius diagram are indeed 50\%~rock-50\%~condensed-water by mass, because most of those exoplanets are too hot for water to be condensed. Our models suggest that an upper equilibrium temperature for the transition would be ~400 K, and other works point to even lower temperatures \citep[e.g.][]{Aguichine21}. The bulk of planets in the sample from \citet{Luque22} have equilibrium temperatures higher than this. 
To analyse this in more detail, comparing the results of our calculations with observations, we plot on the top panels of Fig.\ref{fig_real_synt}, all real exoplanets with $\Teq\leq400$~K for $0.25 \leq\Mstar< 0.55 \, \Msun$ (M-dwarfs) and $\Teq\leq600$~K for $0.55 \leq\Mstar< 0.85 \, \Msun$ (K-dwarfs).

For the M-dwarfs, we note that only 6 real exoplanets have $\Teq\leq400$~K and fairly good mass and radius measurements. Three of those have mass and radius compatible with the condensed water line (within errors). We note that for the planets that have sizes above the condensed-water line, we actually have synthetic planets with that mass and radius (Fig.\ref{fig_real_synt}, bottom-left panel). These planets, despite of having water in condensed form, they have some remnant H/He) that increases the planet radii compared to the condensed-water line.   

For the K-dwarfs (Fig.\ref{fig_real_synt}, right panels), the coldest real planet ($\Teq\approx350$~K) falls exactly on the condensed water-line as our calculations predict. Another 5 real planets fall in between the steam- and condensed-water lines, where our model predicts that the atmospheres should be in the form of steam. Nevertheless, we do have synthetic planets in that region of the parameter space (bottom-right panel). These planets have actually steam atmospheres but the total water content is of $\fw\approx 20-40 \%$ (fig.\ref{Fig_MRfice}). So that could be a plausible explanation for the real exoplanets falling in between the steam- and condensed-water lines for $\Teq\gtrsim400~$K \citet[composition that could also explain the exoplanets analysed by][]{Luque22}.

We also note that while our model yields some steam-worlds with masses between 1 and 3 \ME and $\Teq \approx 500-600$ K for $\Mstar = 0.7~\Msun$, no real exoplanets have been confirmed yet in that part of the parameter space (see top-right panel of Fig.\ref{fig_real_synt}). Nevertheless, the synthetic planets falling in that part of the M-R diagram represent only 5\% of the synthetic water-worlds produced for that stellar mass, so we expect those planets to be rare. In addition, we note that the number of sub-Neptunes around K-dwarfs with $\Teq\leq600$~K is still low to draw strong conclusions.

Another clear difference between the synthetic and real exoplanets observed in this figure is the lack of real rocky exoplanets. This is because of the imposed maximum equilibrium temperature. Real rocky exoplanets exist but typically at higher equilibrium temperatures. Improved modelling on the disc inner edge and the addition of N-body interactions and orbital evolution due to tides is needed to aim at better reproducing the observed distribution of planetary periods.

\end{appendix}

\end{document}